\documentclass{emulateapj}
\usepackage{graphicx}

\slugcomment{Accepted in ApJ}

\newcommand\refsec[1]{\S \ref{sec:#1}}
\newcommand\Rsrc{R_{\rm src}}
\newcommand\Rein{R_{\rm ein}}
\newcommand\Reff{R_{\rm eff}}

\newcommand{\be}{\begin{equation}}
\newcommand{\ee}{\end{equation}}
\newcommand{\ba}{\begin{eqnarray}}
\newcommand{\ea}{\end{eqnarray}}
\newcommand{\bcn}{\begin{center}}
\newcommand{\ecn}{\end{center}}
\newcommand{\ben}{\begin{enumerate}}
\newcommand{\een}{\end{enumerate}}
\newcommand{\btab}{\begin{tabular}}
\newcommand{\etab}{\end{tabular}}
\newcommand{\bt}{\begin{table}}
\newcommand{\et}{\end{table}}
\newcommand{\TWD}{\mbox{TWD}}

\begin{document}

\title{Microlensing of Lensed Supernovae}

\author{Gregory Dobler\altaffilmark{1} \& Charles R. Keeton\altaffilmark{2}}

\altaffiltext{1}{
Department of Physics and Astronomy, University of Pennsylvania,
209 S.\ 33rd Street, Philadelphia, PA 19104 USA
}
\altaffiltext{2}{
Department of Physics and Astronomy, Rutgers University,
136 Frelinghuysen Road, Piscataway, NJ 08854 USA
}

\begin{abstract}
Given the number of recently discovered galaxy-galaxy lens systems,
we anticipate that a gravitationally lensed supernova will be observed
within the next few years.  We explore the possibility that stars in
the lens galaxy will produce observable microlensing fluctuations in
lensed supernova light curves.  For typical parameters, we predict
that $\sim$70\% of lensed SNe will show microlensing fluctuations
$> 0.5$ mag, while $\sim$25\% will have fluctuations $> 1$ mag.
Thus microlensing of lensed supernova will be both ubiquitous and
observable.  Additionally, we show that microlensing fluctuations
will complicate measurements of time delays from multiply imaged 
supernovae: time delays accurate to better than a few days will be
difficult to obtain.  We also consider prospects for extracting
the lens galaxy's stellar mass fraction and mass function from
microlensing fluctuations via a new statistical measure, the
time-weighted light curve derivative.
\end{abstract}


\section{Introduction}

Within the past year there has been an explosion in the number of
known galaxy-galaxy strong lens systems due to the success of the
Sloan Lens ACS (SLACS) survey \citep{bolton04,SLACS1,SLACS2,SLACS3}.  
The lens galaxies consist of a subset of the Sloan Digital Sky
Survey (SDSS) Luminous Red Galaxy (LRG) catalog \citep{LRG} and
were initially selected based on the presence of several ``rogue''
oxygen emission lines in the otherwise well characterized LRG 
spectra \citep{bolton04}.  The hypothesis that these emission 
features originated from an unresolved background galaxy within
the 3\arcsec\ SDSS fiber suggested a substantial probability of
lensing.  Follow-up HST snapshots with the ACS camera revealed
that many of the source galaxies are in fact lensed
\citep{bolton04,SLACS1}.

Oxygen emission lines trace star formation in both the local and
distant universe \citep{kenn,hogg}, so it seems likely that the
SLACS sources are undergoing intense star formation.  Coupled
with the sheer number of lenses discovered with SLACS ($\sim$30
at present count), this opens up the exciting possibility of
observing lensed supernovae (SNe) in the near future.

Lensed SNe offer a variety of interesting applications.  For
example, \citet{og1} showed that SN magnification distributions
can be used to break the degeneracy between the density profile
of the lens galaxy and the measured value of the Hubble constant.
\citet{cooraySN} investigated the possibility of using weakly
lensed SNe to map out large-scale density fluctuations in the
universe.  \citet{kolbar} proposed the use of SN magnification
distributions to eliminate the mass-sheet degeneracy in cluster
mass reconstruction.  In this paper, we point out that strongly
lensed SNe will also be \emph{microlensed} by the stars in the
lens galaxy, and demonstrate that the microlensing fluctuations
will contain valuable information about the stellar population
of the lens galaxy.

The first lensed SN to be discovered will probably be Type II,
as opposed to the Type Ia SNe used to measure the expansion
history of the universe \citep{riess98,perl,tonry03,knop,riess04}.  
While SNe II appear
to be somewhat less common than SNe Ia when averaged over
galaxies of all Hubble types, SNe II have a higher rate among
spiral and irregular galaxies \citep{kenn,mannucci}.  The
presence of [\ion{O}{2}] emission lines in the SLACS sources
suggests that they are predominantly gas-rich spirals and
irregulars.  Since star forming galaxies typically host one
SN II every 50--100 years \citep{mannucci}, we expect a SN II
to be observed in one of the $\sim$30 SLACS source galaxies
within the next several years.  Although there is scatter among
the peak brightnesses of SN II's, the mean B-band luminosity is
approximately $M_B \sim -17.5$ \citep{youngSN}.   This implies
an apparent magnitude of a SN in the SLACS source population
(which has a mean redshift of $z_s \approx 0.53$) of $m \sim 24.5$.
Since there is an additional amplification $\mu_0 \sim 10$ due
to the main lens galaxy, the apparent magnitude for multiply 
imaged SNe in the SLACS sample should be $m \sim 22$, a value
that is certainly within reach of existing telescopes with
reasonable exposure times.

In the SLACS sample, discoveries of lensed SNe will come from
monitoring known lens systems.  The complementary approach is
to mine a catalog of observed SNe to find those that are lensed.
This approach could be used with the Supernova Acceleration Probe
\citep[SNAP;][also http://snap.lbl.gov]{SNAP}, which is intended
to observe $\sim$2000--3000 SNe Ia per year down to 26--27th magnitude.  
\citet{og2} forecast that SNAP will discover 2--20 lensed SNe Ia 
per year, with the uncertainties due mainly to the star formation 
rate at high redshift.  Although SNAP will focus on SNe Ia, the 
deep imaging will naturally reveal many SNe II as well, raising 
the total number of lensed SNe by a factor of $\sim$2.

While there are many categories and subcategories of observed
SNe \citep[see][for a thorough review]{filippenko}, the physics
behind SN explosions has been studied extensively and both light
curves and spectra are reasonably well reproduced in simulations
\citep{eastman2,eastman}.  The progenitors of SNe II are thought
to be massive \citep[$\ga 8$--$10\,M_\odot$;][]{leonard} stars
that undergo core collapse when the amount of energy released in
fusion reactions is no longer sufficient to counteract
self-gravitational forces.  The resulting explosion creates an
expanding envelope that can be modeled with the Expanding
Photosphere Method \citep{kirshner}.  Light curves of some
SN II have been observed with time sampling approaching 1 day
\citep[e.g., SN1999em;][]{leonard}. 

Light that reaches us from a distant SN will of course be affected
by all intervening matter, such as dust and gas, leading to
photometric and spectroscopic differences between the intrinsic
and observed light.  Lensed SNe are unique in that they may also
exhibit gravitationally induced variability due to intervening
stars lying within the lens galaxy.  This phenomenon known as
microlensing is familiar from quasar lensing, with the best
known example being Q2237+030 \citep{wozniak}.  The brightnesses
of the lensed images fluctuate on time scales of months to years
due to relative motion of the lens galaxy and source quasar.  For
a detailed example of microlensing light curve analysis see
\citet{csk}, and for detailed treatments of quasar microlensing
theory see \citet{WT,SW,SWL}.

In a microlensed SN, by contrast, what changes with time is the
\emph{size} of the source.\footnote{The situation is similar
to nanolensing of gamma ray bursts as explored by \citet{walker03}.}
The effects of source size in microlensing have been studied
\citep[e.g.,][]{mortonson,dobler2}, with the general conclusion
that the amplitude of microlensing fluctuations decreases with
increasing source size.  Thus, we na\"ively expect that
microlensing fluctuations may start out large and decrease in
amplitude as the SN expands.  Another qualitative result is
that large sources are not very sensitive to microlensing by
small stars.  Therefore we might hope to probe the mass function
of stars in the lens galaxy by comparing the microlensing at
different times or source sizes.

The notion of using multiple source sizes to ``map out'' lens
mass distributions on various scales has been explored for quasar
lensing by \citet{met} and \citet{dobler}.  In particular,
\citeauthor{dobler} plotted curves of magnification versus source
size and pointed out that multiwavelength observations of quasar
lenses could measure these curves.  However, with quasar lenses
it is possible to obtain only a few discrete points on the curves,
say for X-ray, optical, and radio wavelengths, and there is no
clear mapping between the observable quantity (wavelength) and
source size.  With SNe, by contrast, there is a simple mapping
between the observable quantity (time) and the source size via the
photosphere expansion rate.  Furthermore, with good time sampling
it would be possible to measure the entire magnification versus
source size curve.

Of course, seeing microlensing in SN light curves is contingent
upon having a SN expansion rate that is neither too large (which
would ``wash out'' microlensing effects at early times) nor too
small (which would mean that the SN remains effectively a point
source for its observable lifetime).  In \S 2 we examine the
typical SN expansion rate in units that are meaningful for 
microlensing, and estimate the optical depth for microlensing
by stars in both the lens and host galaxies.  In \S 3 we
describe our microlensing simulations, and in \S 4 we show
sample microlensed SNe light curves.  In \S 5 we interpret the
light curves and consider whether microlensing fluctutations
could be used to extract information about the stellar population:
the relative abundances of stars and smooth dark matter at the
image positions, and the mass function of stars.  We also
consider whether microlensing will affect time delay measurements
in lensed SNe, which have been proposed as a tool for precise
measurements of $H_0$ \citep{og1}.  Finally, in \S 6 we summarize
our findings and discuss possibilities for future theoretical
and observational advances.

\section{Theoretical Background}

\subsection{SN model}

For our microlensing calculations, we model the source SN as an 
expanding disk of uniform brightness, taking as motivation the
success of the Expanding Photosphere Method (EPM) originally
developed by \citet{kirshner}.  EPM assumes that the light
emanating from a SN II explosion is due to a spherically
symmetric, expanding hydrogen shell.  The method (or closely
related variants) has been used by several authors to estimate
distances to observed SNe II \citep{kirshner,schurmann,branch1,branch2}.
The efficacy of EPM rests on two assumptions: the photosphere is
spherical and emits light with a Planck spectrum.  Asphericity 
has been measured to be $<30\%$ via spectropolarimetry 
\citep{leonard}, and departures from Plankian emission due to
flux dilution has been studied numerically
\citep{hershkowitz1,hershkowitz2}.  Lastly, for this pilot study
we also assume that the effects of limb darkening may be neglected.

\subsection{Microlensing scale}

It is important to ask how the size of the SN photosphere compares
to scales relevant for microlensing during the SN's observable
lifetime.  Given a population of stars in the lens galaxy, the
relevant length scale for microlensing is the Einstein radius
\be \label{eq:Rbar}
  \bar{R} = \left( \frac{4 G \bar{m}}{c^2}D \right)^{1/2}
\ee
associated with the mean stellar mass $\bar{m}$.  The distance
factor $D$ is a function of lens and source redshifts ($z_l$ and
$z_s$ respectively) and cosmology.  Throughout this paper we use
$z_l=0.32$ and $z_s=0.53$, which represent the mean lens and source
redshifts of the SLACS sample, and we take $\Omega_M = 0.3$,
$\Omega_{\Lambda} = 0.7$, and $H_0 = 70$ km/s/Mpc.  With these
parameters, equation (\ref{eq:Rbar}) gives
\be
  \bar{R} = 13.1 \left(\frac{\bar{m}}{M_\odot}\right)^{1/2} \mbox{ lt-day} .
\ee
To compare this with the size of the SN, we must convert the
photosphere's expansion rate into units of the Einstein radius.
If the edge of the photosphere is traveling at a velocity $v = 
15,000$ km/s $= 0.05$ c \citep{eastman}, this expansion rate is
\be
  \frac{dR}{dt} = \frac{v}{\bar{R}}\bar{R} 
                = 0.0038 \ \bar{R}\mbox{ day}^{-1},
\ee
implying that the size of the SN is less than or comparable to
$\bar{R}$ for its observable lifetime ($t \sim 300$ days).  In
other words, the SN has a size that should make it quite
sensitive to microlensing.

\subsection{Estimates of self-lensing}

Before examining microlensing by stars in an intervening lens
galaxy, it is interesting to consider whether SNe (both lensed
and unlensed) could be microlensed by stars \emph{in their own
host galaxies}.  While this effect has yet to be identified in
known SNe, the large SN catalogs that will be produced by SNAP
may reveal effects that are rare.  If we imagine that a SN goes
off at the ``back'' (relative to the observer) of its host galaxy,
we can estimate the self-lensing optical depth\footnote{Following
the standard convention \citep{vo83,pac86,krs86,WT}, we define
the microlensing optical depth to be the fraction of the image
plane area covered by the Einstein disks of stars.  This is
equal to the convergence $\kappa_*$, or the surface mass density
of stars in units of the critical density for lensing.} by
integrating along the line of sight through the host galaxy,
\be \label{eq:selflensingkappa}
  \kappa_* = \int_{0}^{2L} n\,\pi\,R^2\,d\ell\,,
\ee
where $2 L$ is the diameter of the host galaxy, $n$ is the number
density of stars, and $R$ is again the Einstein radius of an
individual star.  To make a simple toy model, we assume that
$n$ is constant through the galaxy, that all of the stars have
the same mass $m$, and that the galaxy can be approximated as a
sphere of radius $L$.

From the definition of the Einstein radius, equation 
(\ref{eq:selflensingkappa}) can be written as
\be
  \kappa_* = \int_{0}^{2L} n\ \frac{4 \pi G m}{c^2}\ 
    \frac{d_l d_{ls}}{d_s}\ d\ell\,,
\ee
where $d_l$ is the distance to the lens, $d_s$ is the distance
to the source, and $d_{ls}$ is separation between the lens and
the source.  Since the distance to the host galaxy will be much
larger than the galaxy itself, $d_s \approx d_l$ and $d_{ls} = \ell$.
Integrating then yields
\be
  \kappa_* \sim \frac{6 G m_{gal}}{c^2 L}\ ,
\ee
where we have used the fact that $n m = \rho$ is the mass
density of the galaxy, and we have assumed a uniform spherical
galaxy so $\rho \sim (3 m_{gal})/(4 \pi L^3)$ where $m_{gal}$ is
the total \emph{stellar} mass in the galaxy.  Note that the
self-lensing optical depth is independent of the distance to
the host galaxy.  Using typical values for a Milky Way type
galaxy, $m_{gal} \sim 10^{11} M_{\odot}$ and $L \sim 15$ kpc,
we find
\be
  \kappa_* \sim 2 \times 10^{-6} .
\ee
We conclude that the optical depth is too small, even for a SN
at the back of the galaxy, to produce a significant number of
self-lensing events even in the large catalogs that will be
produced by SNAP.

\subsection{Stellar mass fraction}
\label{sec:estfstar}

The microlensing optical depth is characterized by the convergence
$\kappa_*$, or the surface mass density of stars in units of the
critical density for lensing \citep[see][]{vo83,pac86,krs86,WT}.
It is convenient to define the fraction of the surface density in
stars, $f_* = \kappa_*/\kappa_{\rm tot}$, at the image positions.
To make a rough estimate of $f_*$, we model the lens galaxy's total
mass distribution as a Singular Isothermal Sphere
\citep[see][]{SLACS3},
\be
  \kappa_{SIS} = \frac{\Rein}{2r},
\ee
where $\Rein$ is the Einstein radius.  We model the stellar
mass distribution with a de Vaucouleurs model,
\be
  \kappa_{DeV} = A\,e^{-k\left(r/\Reff\right)^{1/4}},
\ee
where $\Reff$ is the effective radius, and $k = 7.67$.  If we
set the stellar mass-to-light ratio equal to its maximum possible
value such that $\kappa_{DeV} \not> \kappa_{SIS}$ for all $r$, we
can find the maximum value of the normalization constant $A$.
With this maximum value of $A$ we can obtain an upper bound on
$f_{*} = \kappa_{DeV}/\kappa_{SIS}$ as a function of radius.

\begin{figure}
\begin{center}
\includegraphics[width=3.2in]{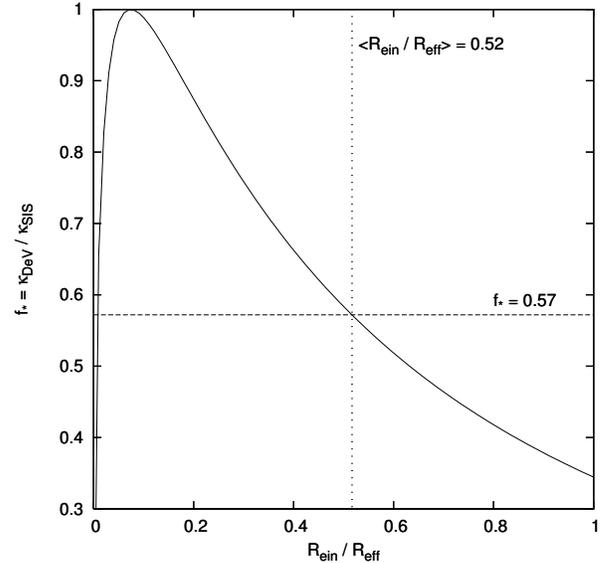}
\end{center}
\caption{
Maximum $f_*$ as a function of $\Rein/\Reff$ assuming a de Vaucouleurs 
profile for the light and an isothermal profile for the mass.  
\citet{SLACS3} measure $\langle \Rein/\Reff \rangle = 0.52$ for the
SLACS lenses.  The maximum surface density in stars at that radius
is $f_* = 0.57$.
}\label{fig:kappa-us}
\end{figure}

Figure \ref{fig:kappa-us} shows the upper limit on $f_{*}$ at the
Einstein radius, as a function of $\Rein/\Reff$.  The average
Einstein radius of the SLACS lenses is
$\langle \Rein/\Reff \rangle = 0.52$ \citep{SLACS3}, and this
corresponds to a maximum stellar mass fraction of $f_{*} = 0.57$.
Also interesting to consider is the integrated stellar mass
fraction within the Einstein radius.  We find the maximum value
of this integrated stellar mass fraction to be 0.83 for
$\langle \Rein/\Reff \rangle = 0.52$.  For comparison,
\citet{SLACS3} find the mean integrated stellar mass fraction to
be 0.75 for the SLACS lenses, suggesting that that the actuall
stellar mass density is $\sim$90\% of its maximum possible value.
This would imply that $f_{*} = 0.46$ at the Einstein radius
(again for $\langle \Rein/\Reff \rangle = 0.52$).  Therefore,
we consider $f_{*} \approx 0.5$ to be a reasonable estimate of
the stellar mass fraction near the image locations of the SLACS
lenses

\section{Microlensing Simulations}
\label{sec:MLsims}

\begin{figure*}
\centerline{
  \includegraphics[width=2.8in]{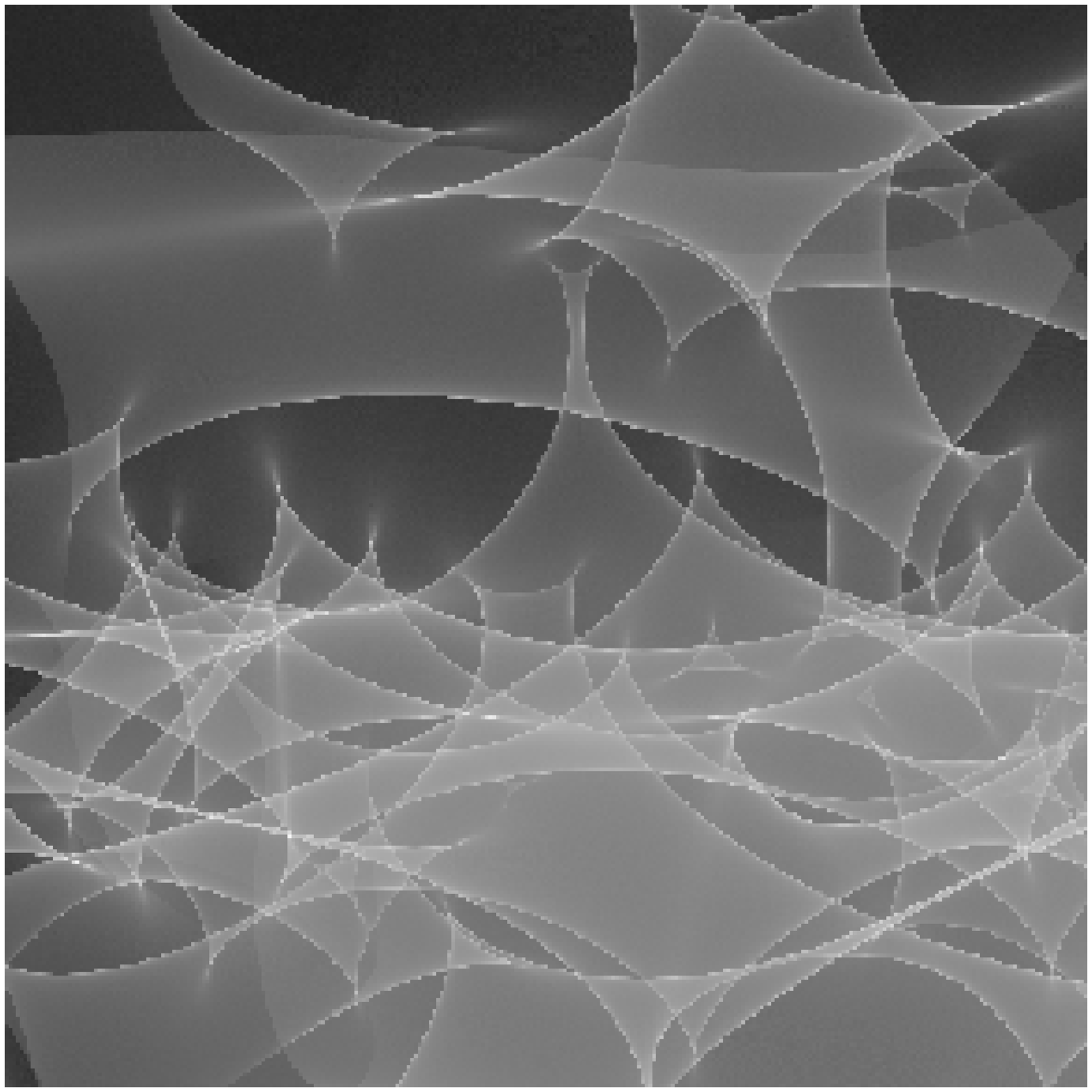}
  \includegraphics[width=2.8in]{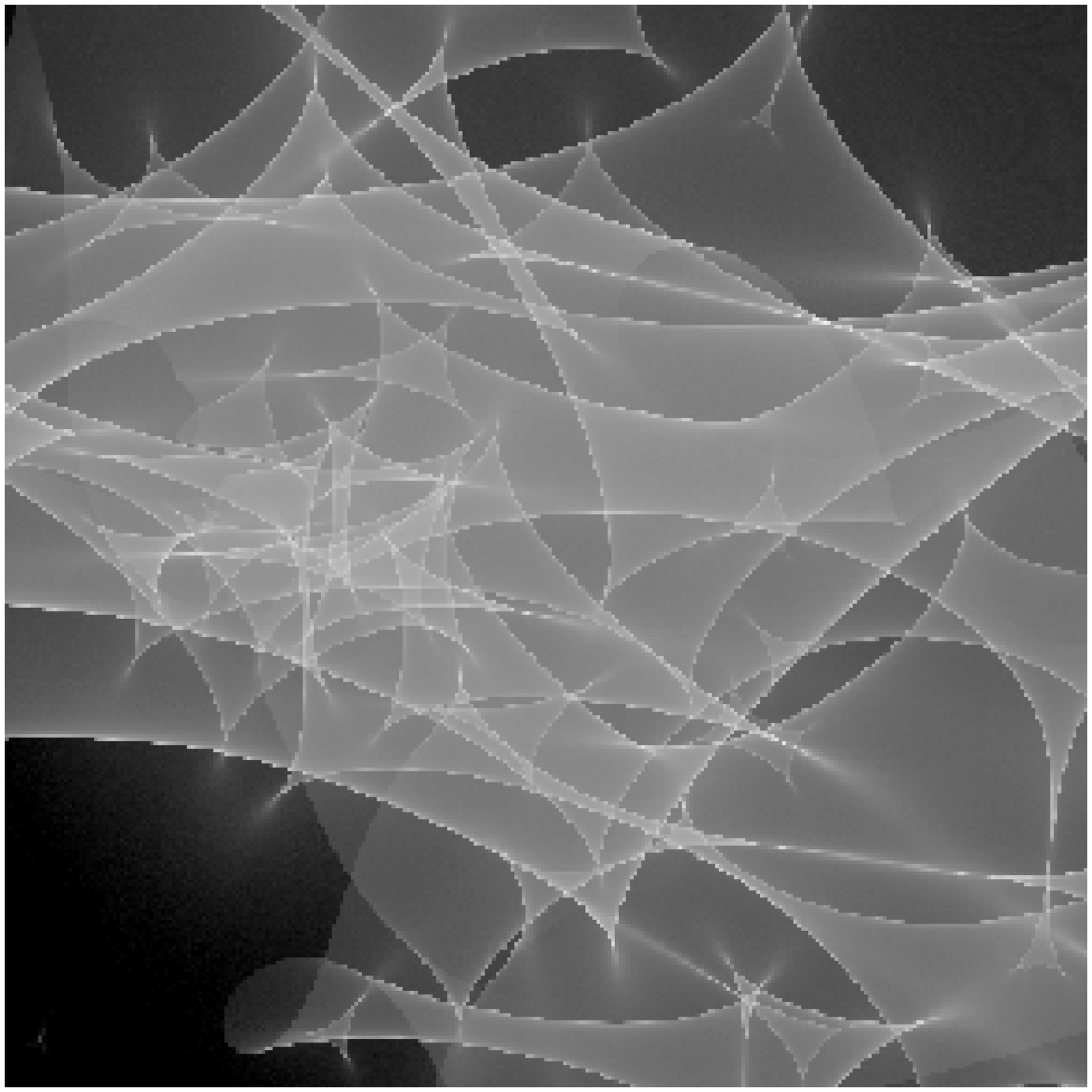}
}
\centerline{
  \includegraphics[width=2.8in]{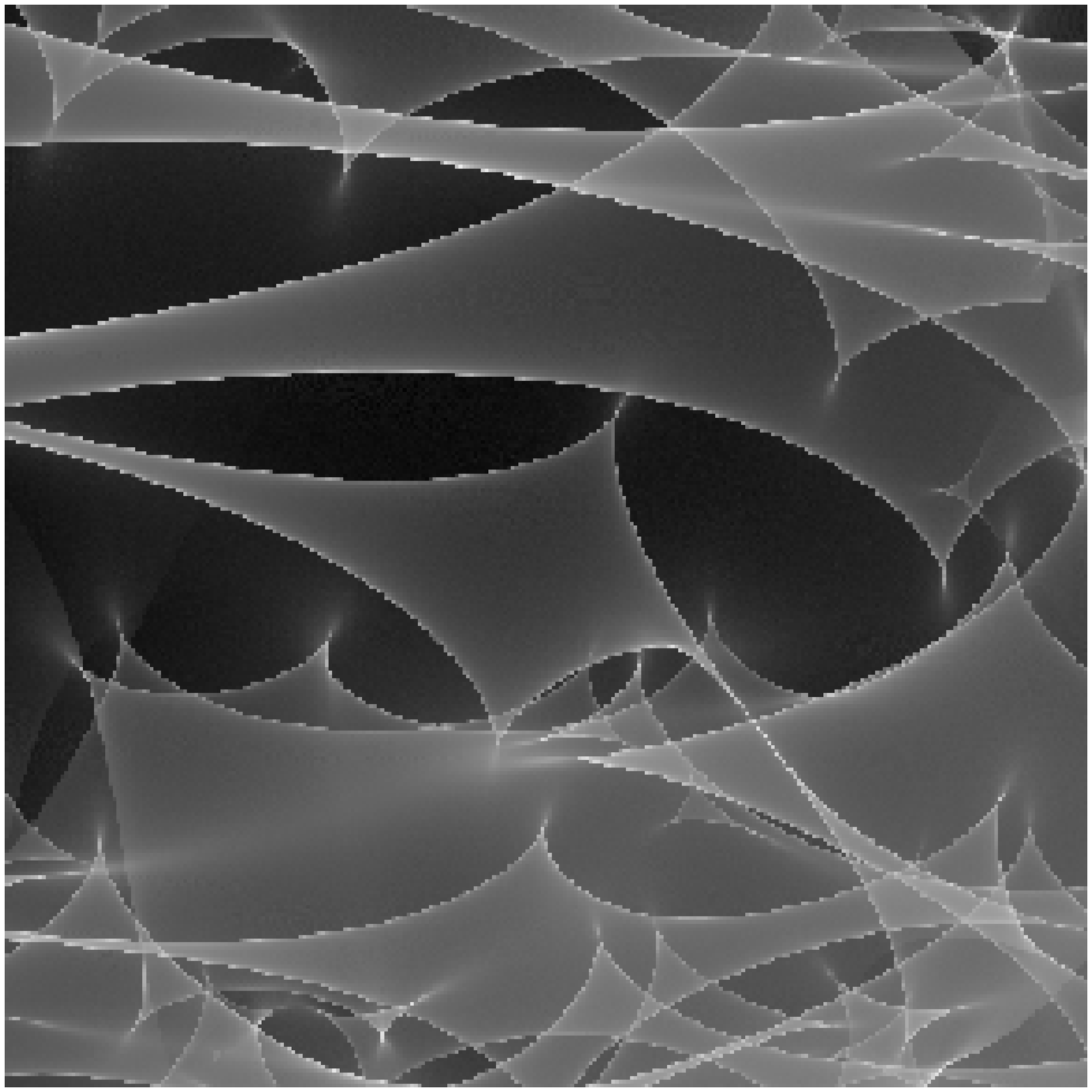}
  \includegraphics[width=2.8in]{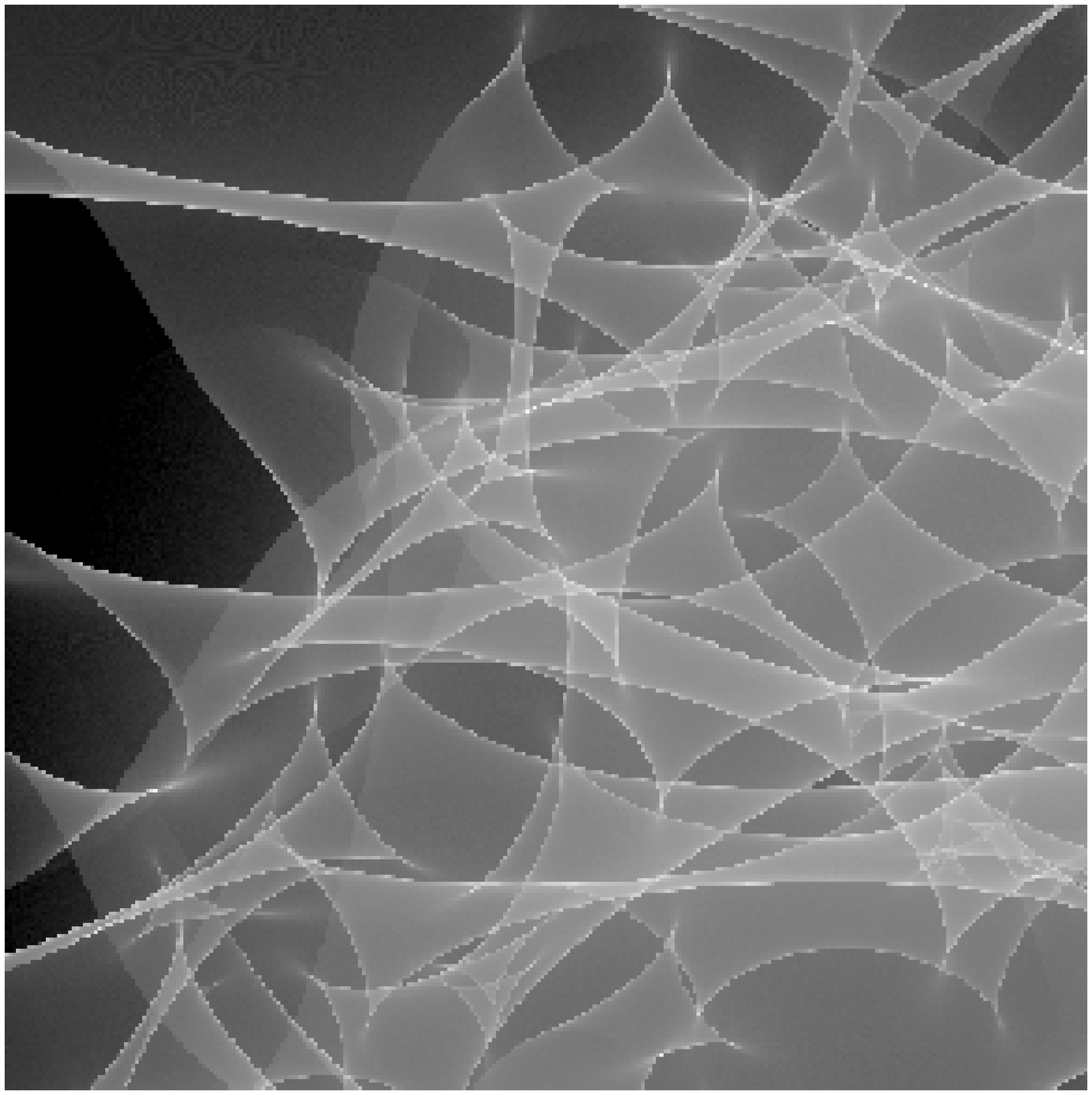}
}
\caption{
Sample magnification maps for four different realizations of stellar
populations with $f_{*} = q = 1.0$ and $\bar{m}=M_{\odot}$.  (These
maps were generated with the code by \citet{jcam}, although we
actually use our own microlensing code to compute SN light curves.)
The side length of each map is $2.5 \bar{R}$, so an expanding SN
will grow to cover approximately one half of a map.  The large
qualitative differences in the magnification maps imply that
microlensing fluctuations will be very different from one SN light
curve to the next, even for fixed stellar population parameters.
}\label{fig:magmapsSN}
\end{figure*}

We model the population of stars using a Salpeter mass function,
$dn/dm \propto m^{-2.35}$, spanning the mass range from $m_1$ to
$m_2$.  We find it convenient to specify the mass range by
setting the mean mass to be $\bar{m} = M_\odot$ and quantifying
the dynamic range with $q = m_1/m_2$.  Several authors have shown
that microlensing magnification distributions are insensitive to
the mean mass $\bar{m}$ \citep{WT,SWL,mortonson}.  For microlensing
of SN light curves, the mean mass does set an overall time scale:
as $\bar{m}$ increases, the time it takes for the source to reach
a given fraction of $\bar{R}$ scales as $\bar{m}^{1/2}$.

The properties of the microlensing fluctuations should be dependent
on the stellar population parameters (namely $f_{*}$ and $q$).
They also depend on the particular configuration of stars, because
as shown in Figure \ref{fig:magmapsSN} different realizations of
stellar populations that are statistically equivalent yield very
different magnification maps in the source plane.  Note that these
magnification maps are $2.5 \bar{R}$ on a side, so a SN will expand
to cover roughly half of a map during its observable lifetime.

We compute microlensing magnifications using a new ray-shooting
code.  The code uses a simplified version of a tree algorithm to
efficiently compute deflection angles for a population of stars.
Whereas the ray-shooting algorithm used by \citet{jcam} is designed
to provide magnifications for many source positions (yielding
magnification maps), our new algorithm is optimized for handling
a wide range of source sizes.  Details of the code will be
presented elsewhere (Dobler \& Keeton, in prep.).

\section{Results}
\label{sec:results}

In our first set of simulations we set $\kappa = \gamma = 0.45$,
where $\kappa = \kappa_* + \kappa_s$ is the total surface mass
density, $\kappa_s$ is the surface density of a smooth dark matter
component, and $\gamma$ is the external shear.  We choose
$\kappa = \gamma = 0.45$ for illustrative purposes as these are
typical values for a positive parity image near the Einstein radius
of an isothermal lens galaxy.  In the absence of microlensing
(i.e., if there were only ``macrolensing'' due to the lens galaxy),
this would yield a magnification of $\mu_0 = 10$.

\begin{figure}
\begin{center}
\includegraphics[width=3.2in]{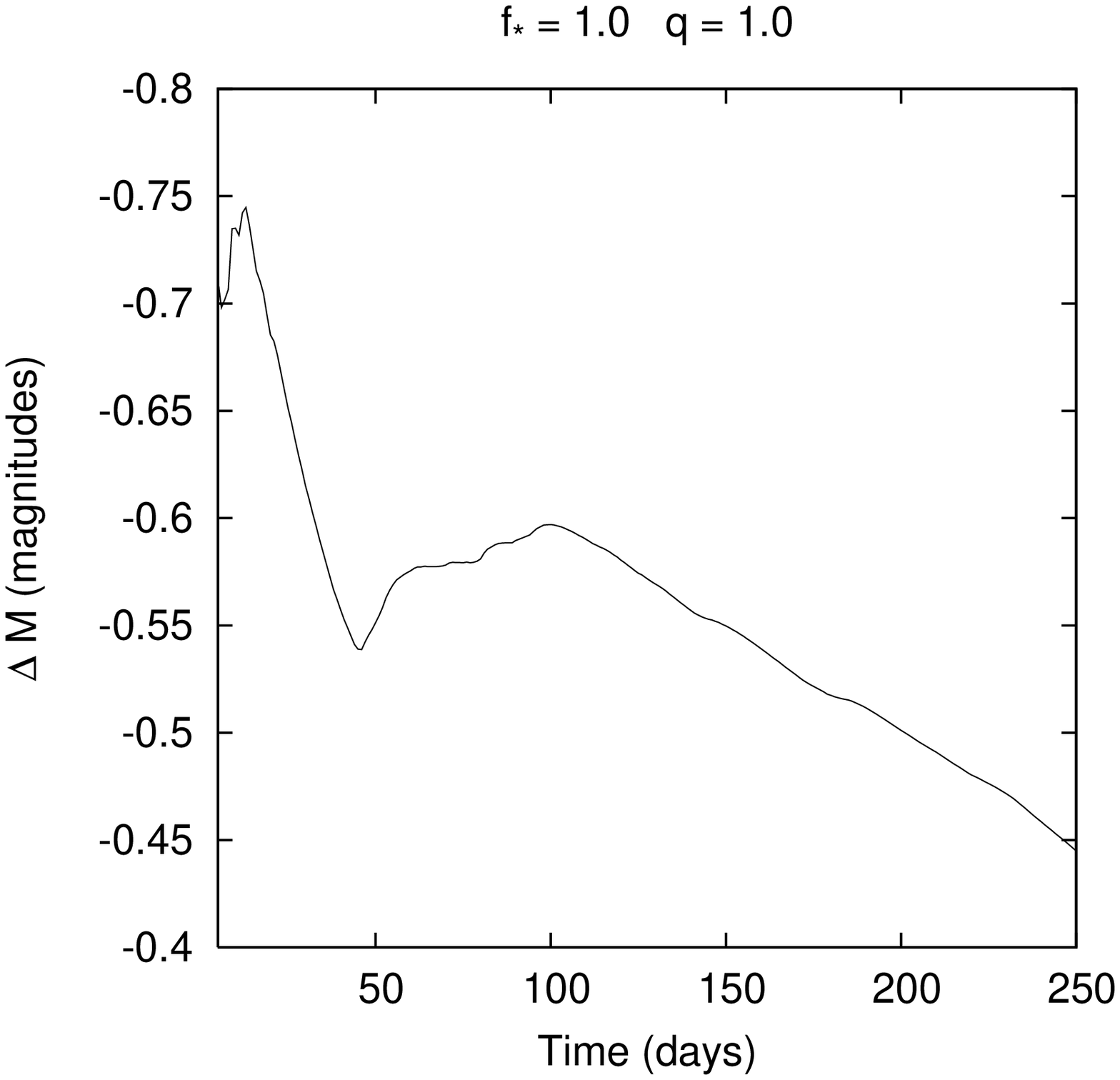}
\end{center}
\caption{
A sample difference light curve for SN microlensing with a uniform mass 
function ($q=1$) and 100\% of the surface density in stars ($f_*=1$).
Microlensing produces offsets of nearly 1 mag, and shorter time scale
fluctuations as large as 0.1--0.2 mag.
}\label{fig:excurves}
\end{figure}

Initially we isolate microlensing fluctuations by plotting the
\emph{difference light curve},
\be
  \Delta M(t) = -2.5 \log\left[\frac{I(t)}{\mu_0 S(t)}\right] ,
\ee
where $S(t)$ and $I(t)$ are the intrinsic and observed brightnesses
of the SN.  This definition is chosen so that $\Delta M = 0$ in
the absence of microlensing.  In general we expect
$\Delta M(t) \to 0$ as $t \to \infty$ since large sources are
insensitive to microlensing.  Figure \ref{fig:excurves} shows a
sample difference ligth curve for one stellar realization with a 
uniform mass function ($q=1$) and 100\% of the surface density in
stars ($f_*=1$).  Over the observable lifetime of the SN, this
stellar configuration creates an \emph{offset} of up to 0.75 mag,
and \emph{fluctuations} at the level of 0.1--0.2 mag.  These effects
are well within the realm of detectability.

\begin{figure*}
\centerline{
  \includegraphics[width=2.8in]{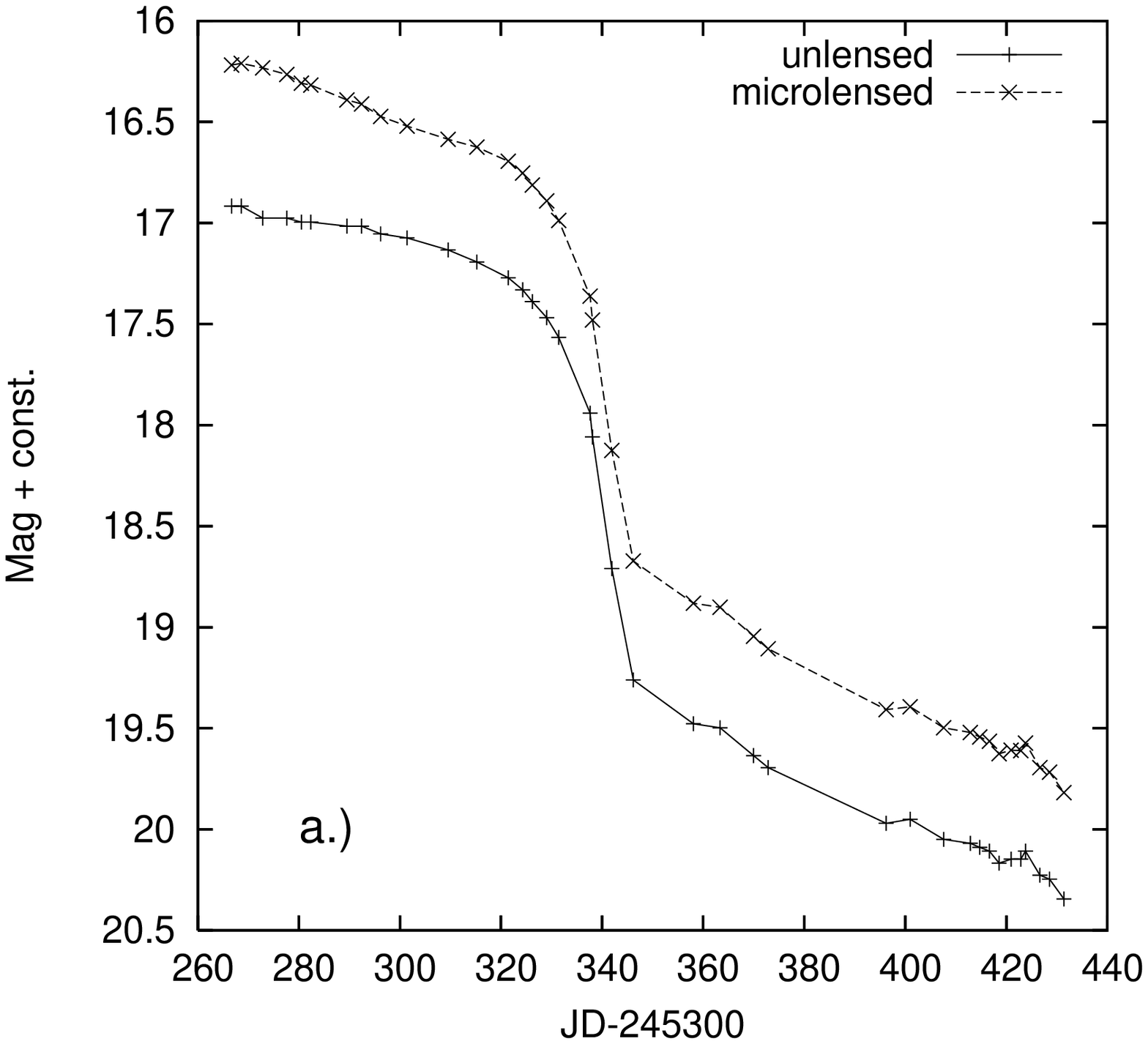}
  \includegraphics[width=2.8in]{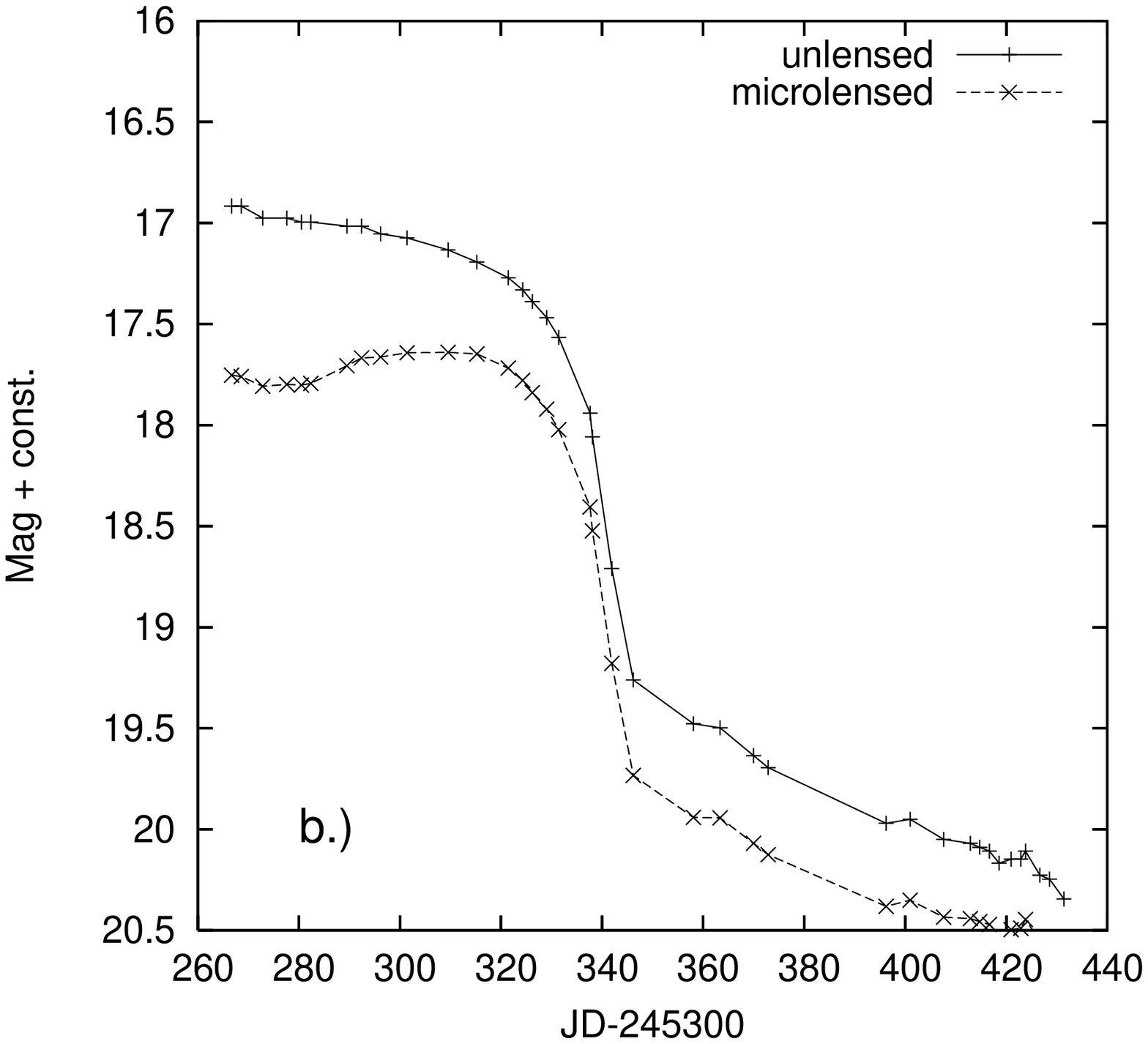}
}
\centerline{
  \includegraphics[width=2.8in]{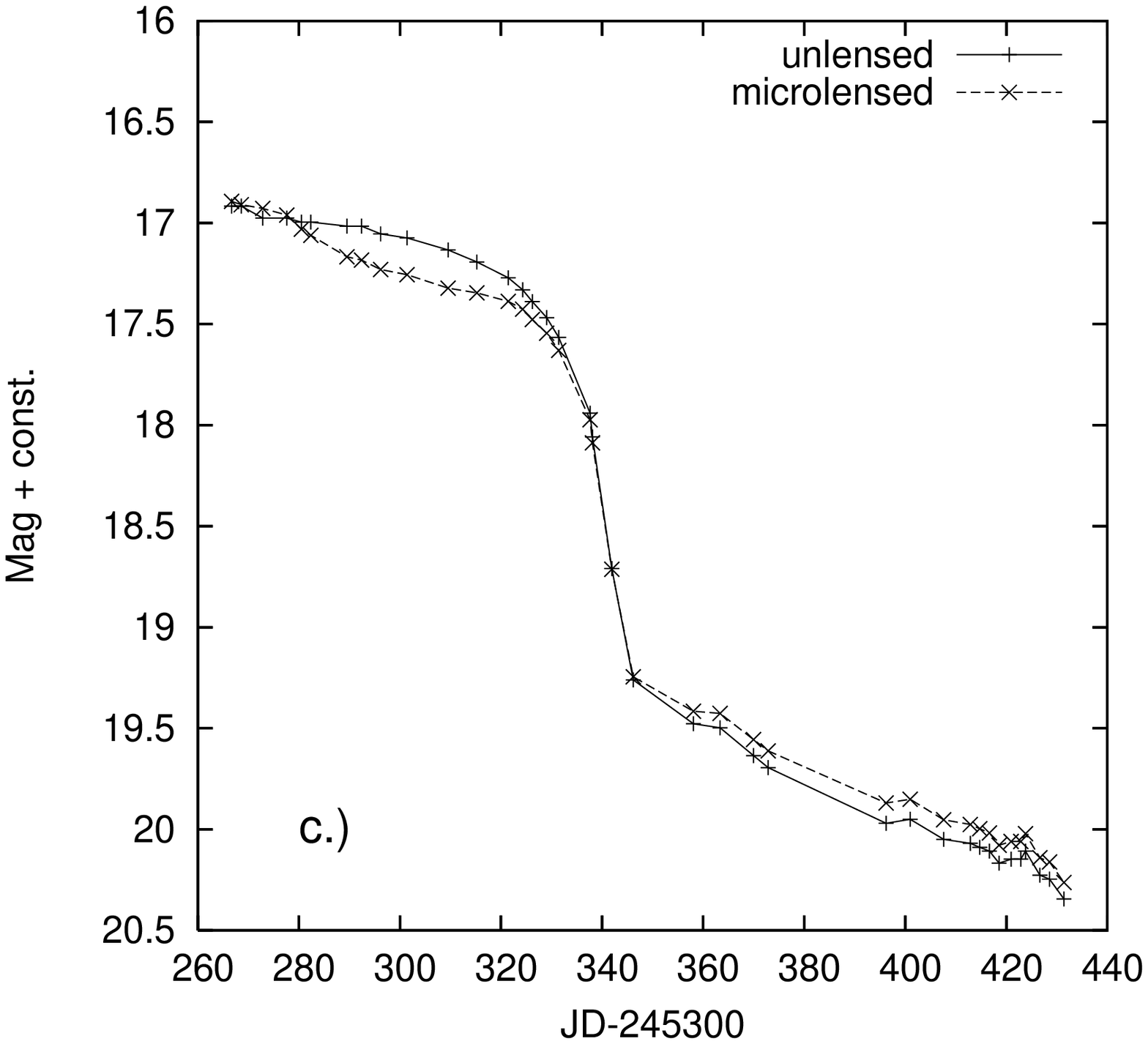}
  \includegraphics[width=2.8in]{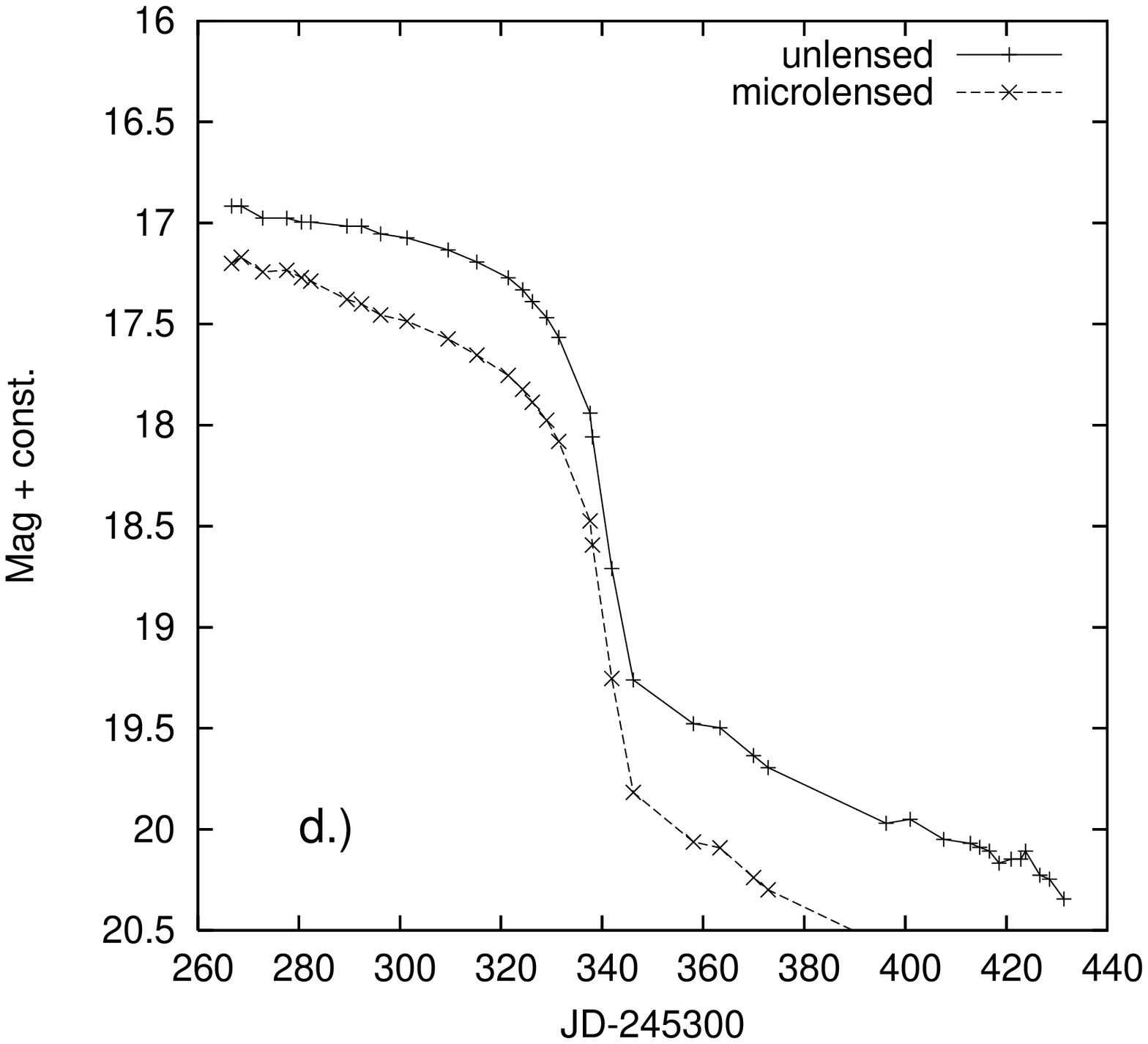}
}
  \caption{
Sample light curves for SN 2004ej, both with (dashed) and without
(solid) microlensing fluctuations.  The intrinsic light curve data
are taken from the Carnegie Supernova Project.  The four panels
represent different stellar configurations which lead to numerous
qualitative changes in the observed light curve.  Microlensing can
produce (a) a large offset, (b) non-intrinsic features such as
peaks, (c) small fluctuations, or (d) changes to the decay rate.
}\label{fig:datafluc}
\end{figure*}

We now combine difference light curves with data for the observed
SN 2004ej from the Carnegie Supernova Project
(CSP)\footnote{http://csp1.lco.cl/\~{}cspuser1/CSP.html}.
SN 2004ej is a classic example of a SN II-P (plateau) in which
the light curve falls off slowly at early times, drops suddenly
by almost 2 mag, and then decreases slowly thereafter.
Figure \ref{fig:datafluc} shows four examples of microlensing
effects.  Depending on the specific stellar configuration, the
SN light curve can be changed either subtly or dramatically.  In
particular, panel (a) shows a large offset due to microlensing,
plus small-amplitude fluctuations.  Panel (b) shows not only an 
offset of $\sim$1 mag, but also a pronounced peak that is not 
present in the intrinsic light curve.  The ability of microlensing
to induce such large-scale features may limit the precision with
which one can ``match up'' light curves from different lensed
images to measure precise lensing time delays (see \refsec{dt}).
Panel (c) shows that the microlensing fluctuations can be more
subtle with only small fluctuations about the intrinsic light
curve, while panel (d) shows that microlensing can substantially
alter the decay rate of the SN light curve.

\begin{figure*}
\centerline{
  \includegraphics[width=2.0in]{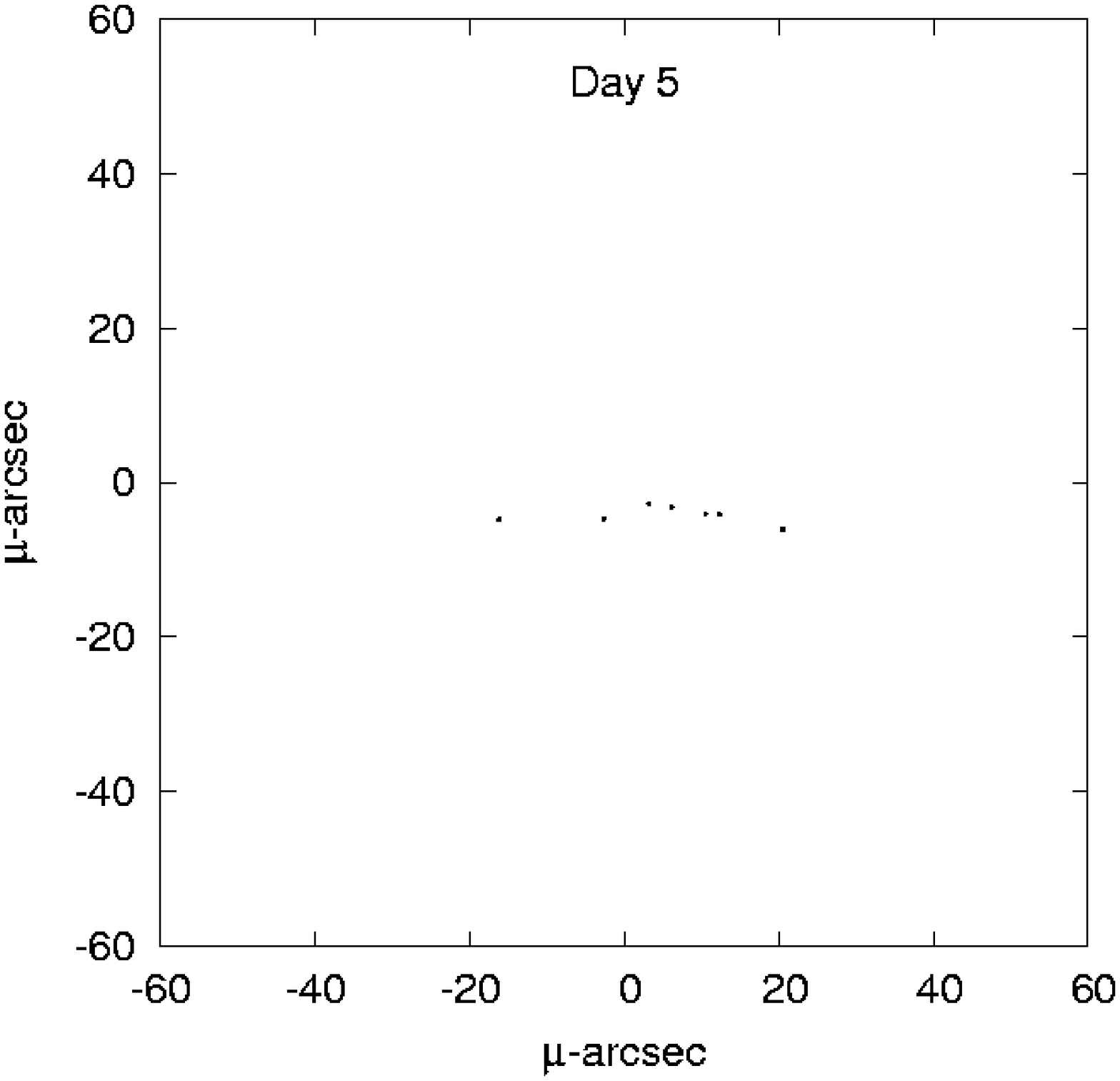}
  \includegraphics[width=2.0in]{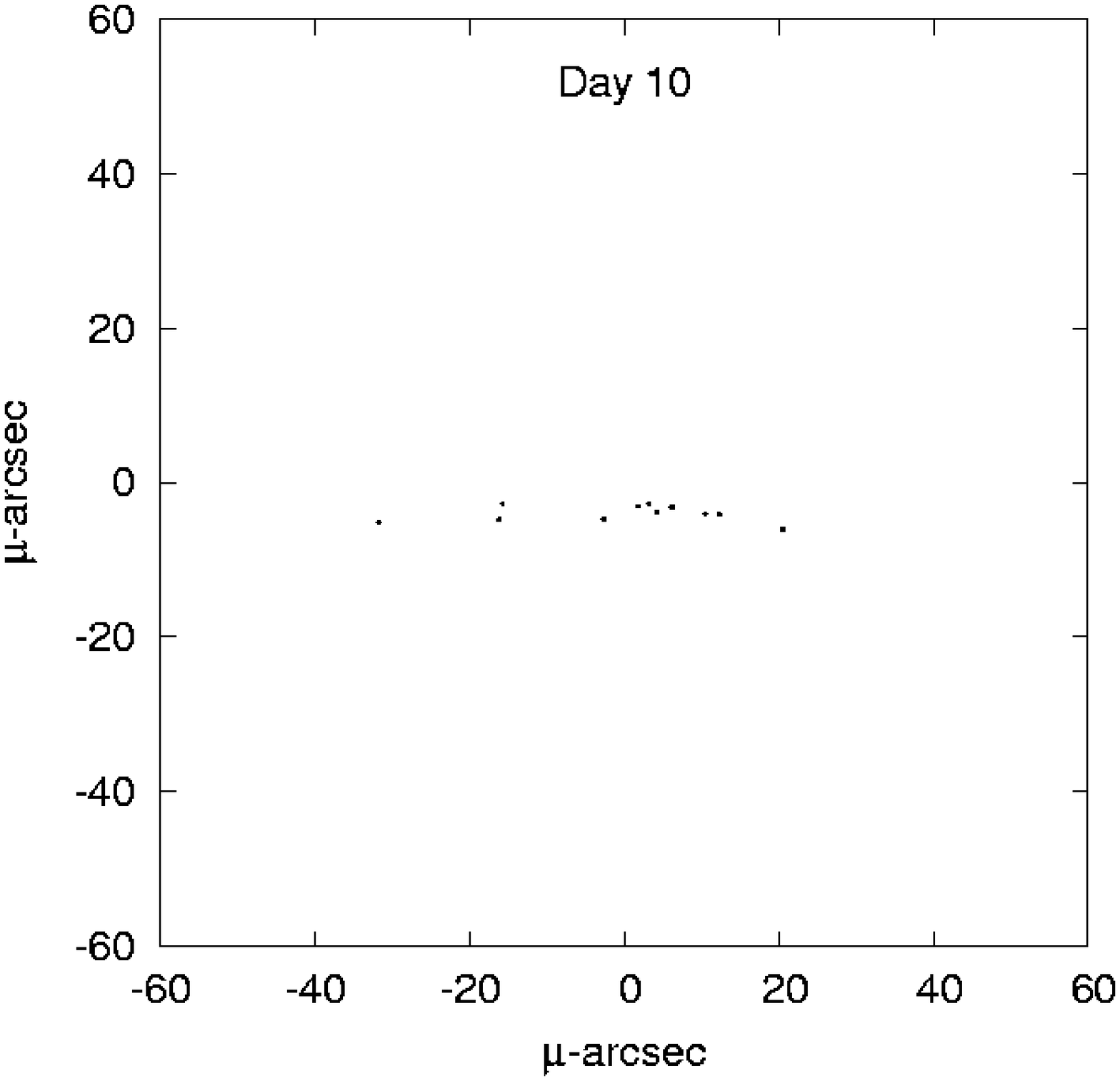}
  \includegraphics[width=2.0in]{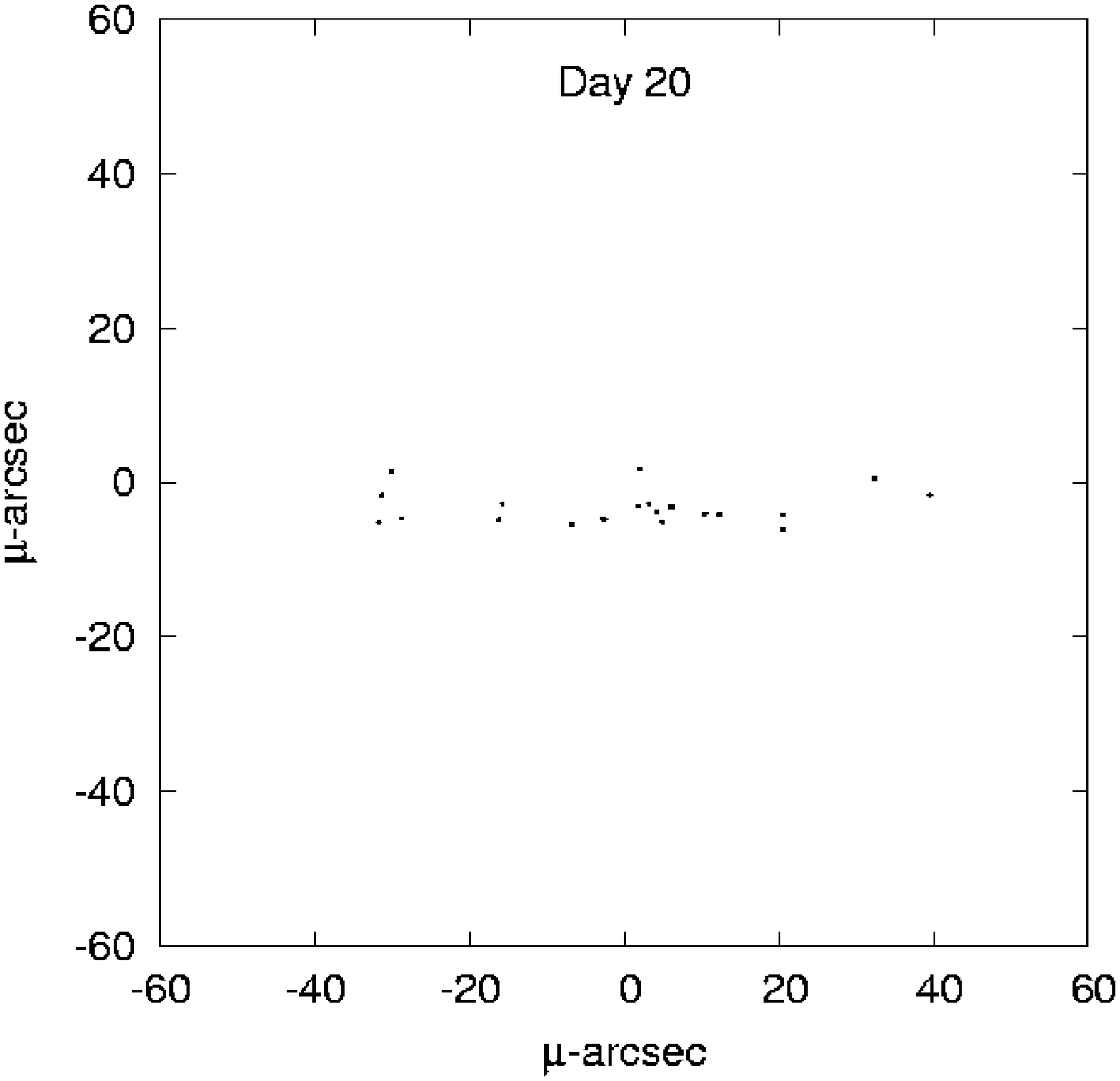}
}
\centerline{
  \includegraphics[width=2.0in]{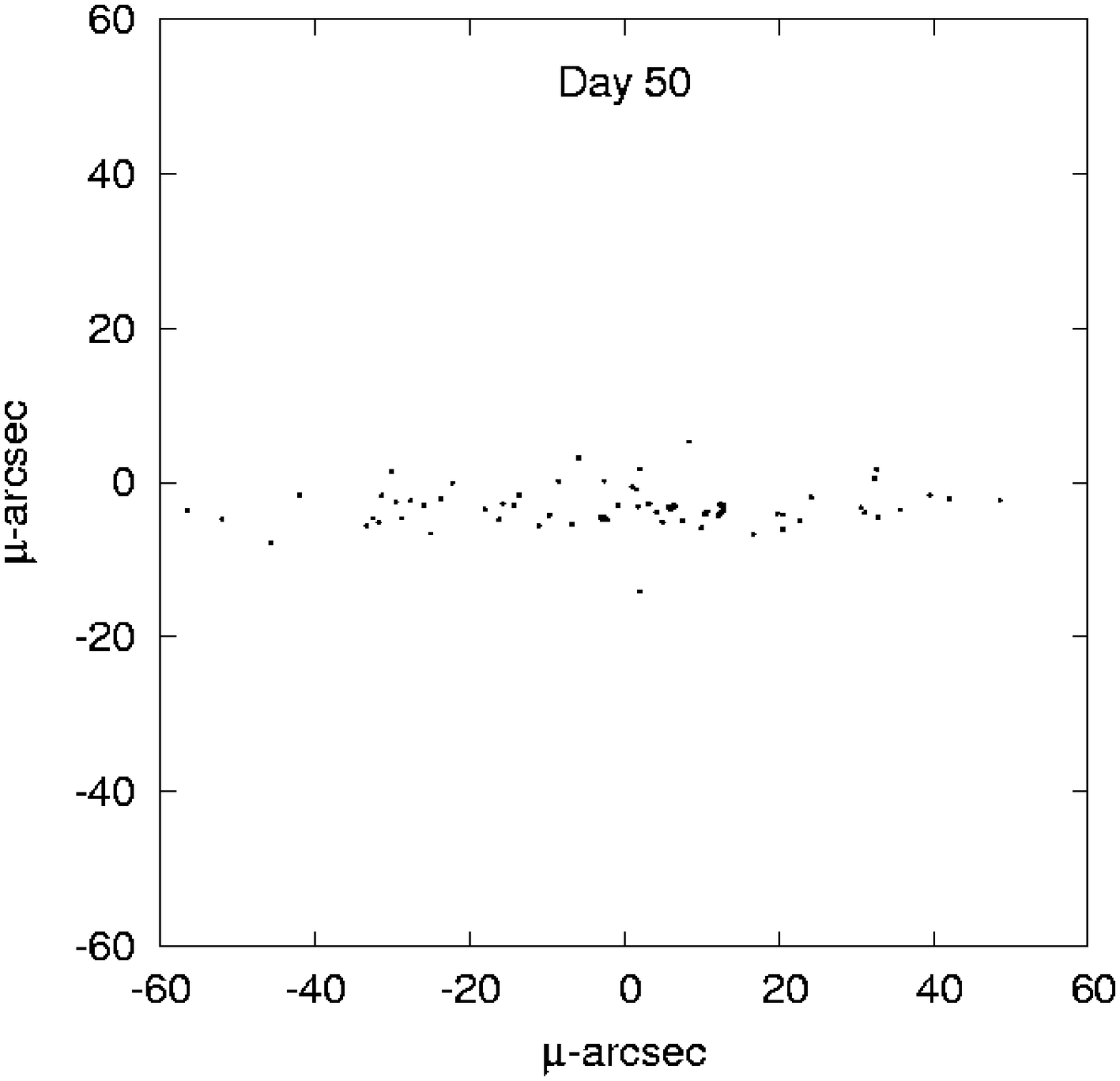}
  \includegraphics[width=2.0in]{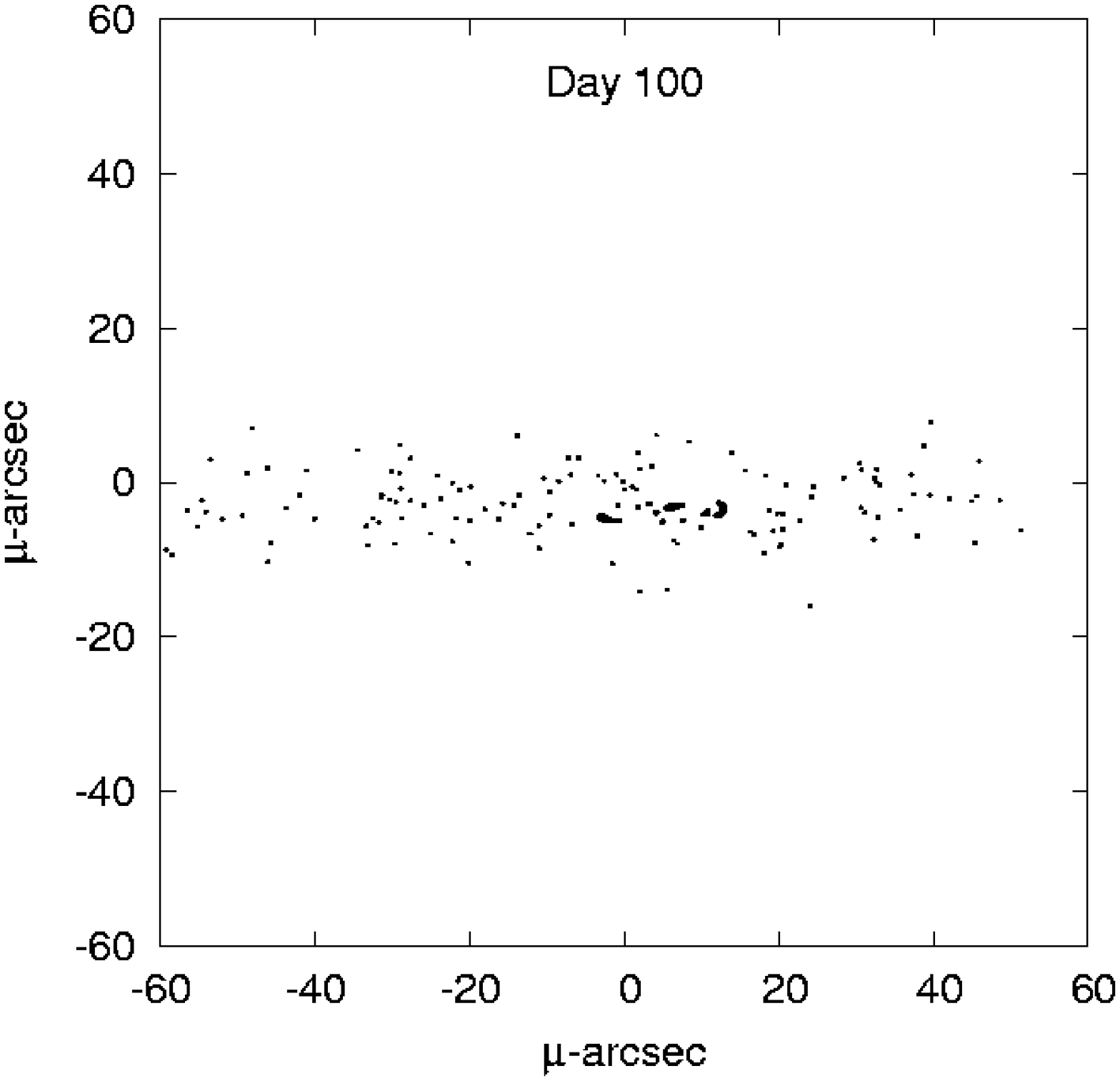}
  \includegraphics[width=2.0in]{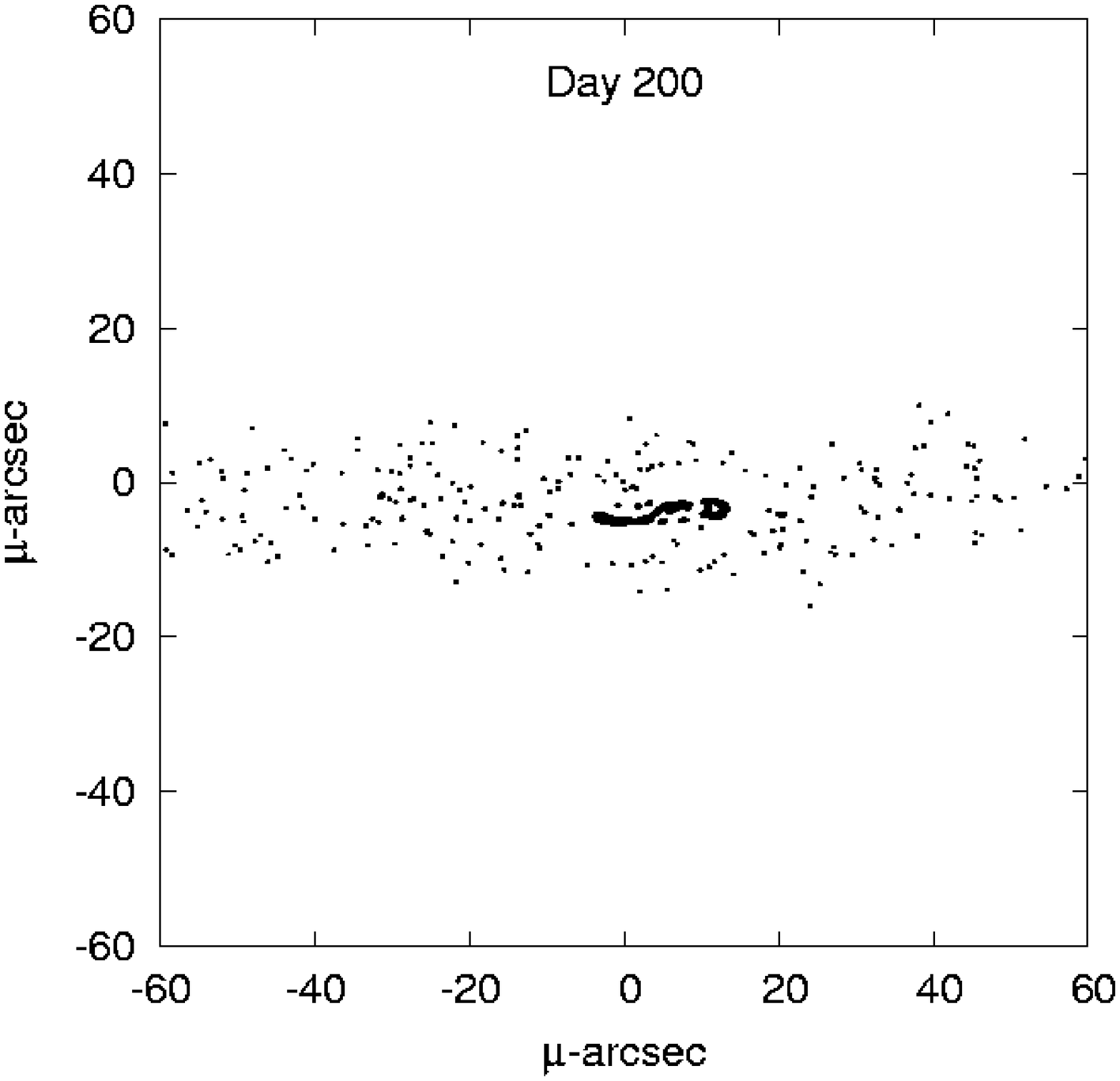}
}
\caption{
Micro-image configurations at various epochs for $q=f_*=1.0$.  As
the SN grows in size, micro-images appear and merge.  The source
size of the photosphere for these image configurations is
$\Rsrc =$ 0.019, 0.038, 0.076, 0.19, 0.38, and 0.76 $\Rein$ for
times $t =$ 5, 10, 20, 50, 100, and 200 days respectively.
The photometric effects on the intrinsic SN light curve for this
particular stellar configuration are shown in panel (b) of
Figure \ref{fig:datafluc}.
}\label{fig:imgconfig}
\end{figure*}

The lensed image configurations for various epochs of panel (b)
are shown in Figure \ref{fig:imgconfig}.  Even when the SN is
small there are multiple micro-images, and no single image is
dominant.  As the SN expands, more and more small micro-images
appear, and a few larger micro-images near the center begin to
merge.  The distribution of micro-images is elongated along the
shear axis (the horizontal axis in Fig.~\ref{fig:imgconfig}).
Even at $t = 200$ days ($\Rsrc = 0.76 \Rein$), by which time
several merged images have formed, there is still no clear
``primary'' micro-image.  The complexity of the image
configurations is striking and attests the richness of the
phenomenon.

\begin{figure}
\begin{center}
\includegraphics[width=3.2in]{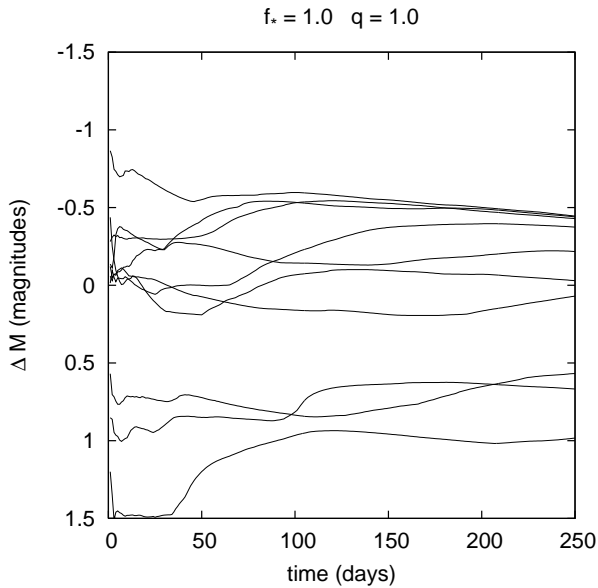}
\end{center}
\caption{
Difference light curves for fixed $f_* = q = 1.0$ but varying
stellar configurations.  Short time scale fluctuations $\sim$0.2 mag
are quite common, while longer time scale variations can be as
large as 0.5--1.0 mag.
}\label{fig:exall-lcs}
\end{figure}

\begin{figure*}
\begin{center}
\includegraphics[width=5.5in]{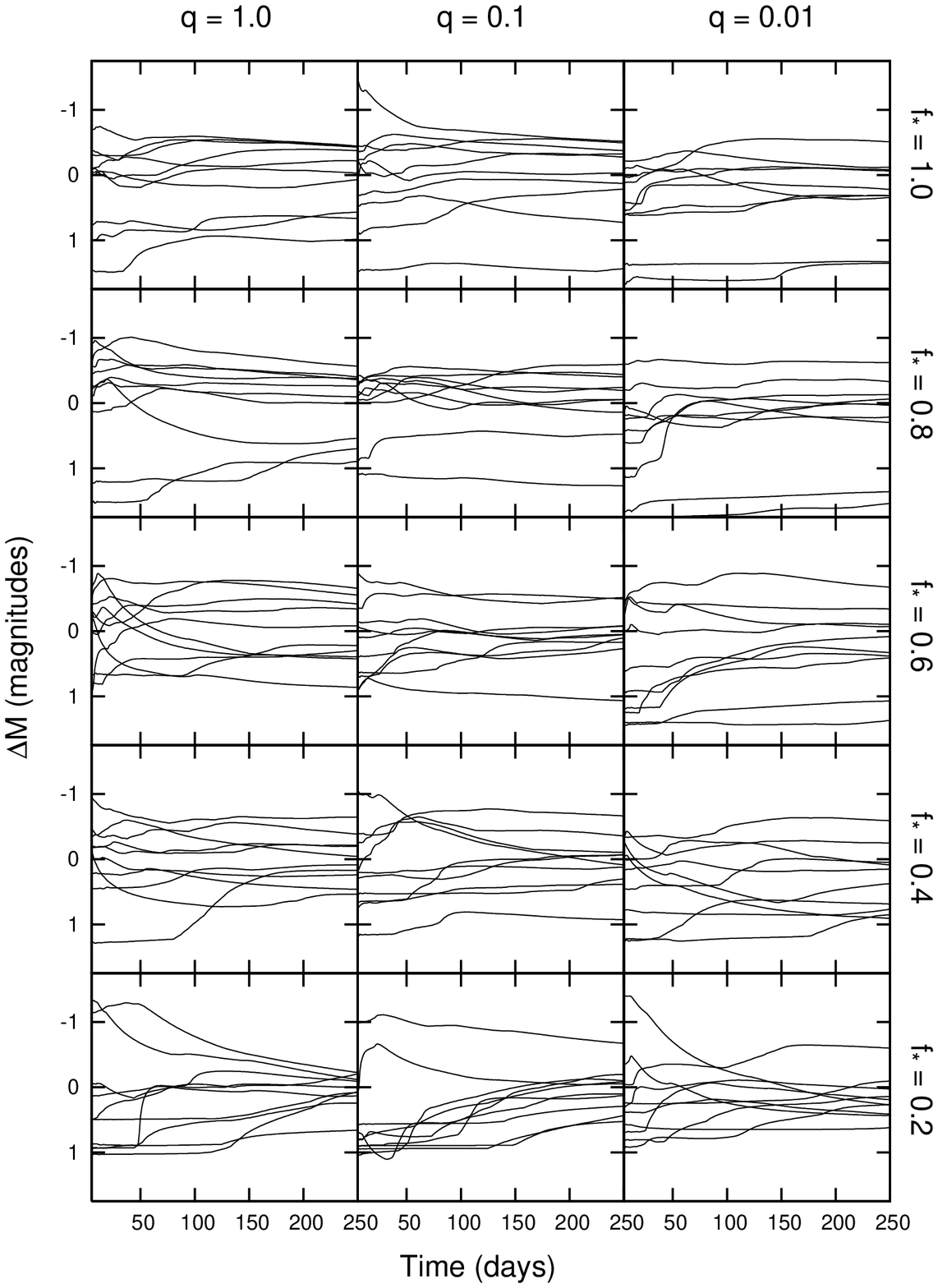}
\end{center}
\caption{
Difference light curves for various $q$ and $f_*$.  Ten realizations
of the stellar configuration are shown for each case.  Significant
microlensing fluctuations are seen in every realization and for every
combination of $q$ and $f_*$, implying that microlensing will be quite 
common in lensed SNe.
}\label{fig:fullsample}
\end{figure*}

Figure \ref{fig:exall-lcs} shows difference light curves for various 
random star populations (all with $q = f_* = 1$).  It is clear that
\emph{all} of the realizations produce offsets, which can be as large
as 1--1.5 mag.  The amplitude of the smaller scale fluctuations
depends on the timescale.  That is, there are occasionally very
rapid variations at the level of $\sim$0.2 mag, while longer time
scale variations can be as large as $\sim$0.5--1.0 mag.  To test
if these trends are specific to these particular stellar population
parameters, Figure \ref{fig:fullsample} shows the effects of varying
$q$ and $f_*$.  We show 10 realizations of the stellar population
for each parameter combination.  It is clear that microlensing
fluctuations are very generic.  While at first glance it is not
obvious whether the realizations differ substantially from one
panel to another, each \emph{individual} curve has an enormous
amount of structure.  It would seem that a lot of information
about the stellar population is hidden in the light curves; the
challenge is figuring out how to extract it (see \refsec{extract}).

\begin{figure*}
\centerline{
  \includegraphics[width=3.0in]{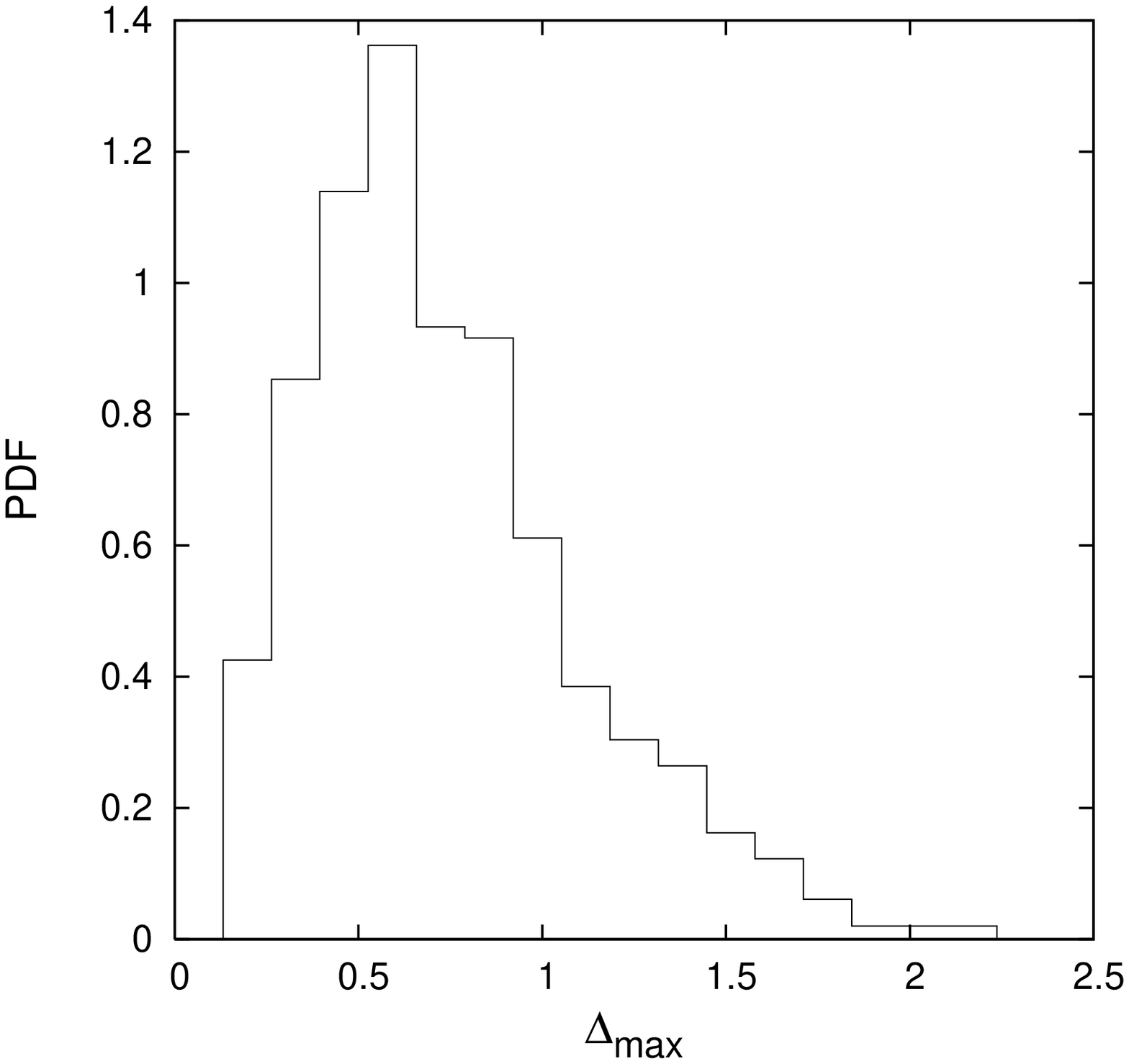}
  \includegraphics[width=3.0in]{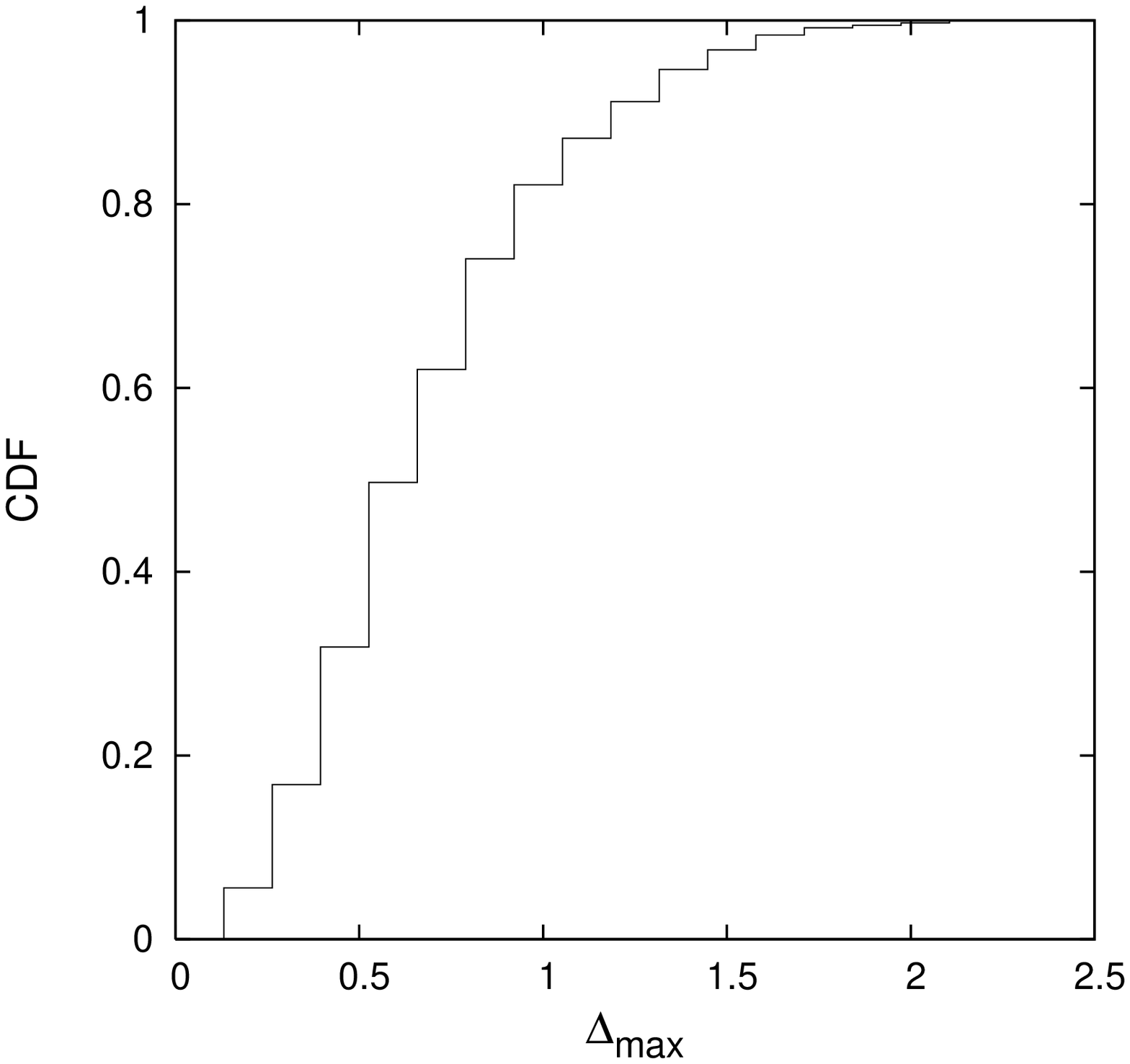}
}
\caption{
The probability density function (left) and cumulative distribution
function (right) for $\Delta_{\rm max}$, averaged over all stellar 
population parameters and configurations.  The PDF is peaked at 
0.6 mag with a long tail extending to $> 2$ mag.  The CDF shows that
roughly 70\% of observed SN light curves will exhibit microlensing
fluctuations $> 0.5$ mag, and 25\% will show fluctuations $> 1$ mag.
}\label{fig:delmaxPDF}
\end{figure*}

Figure \ref{fig:fullsample} also shows that there is not a single 
realization that does \emph{not} produce fluctuations over the
lifetime of the SN.  To quantify this point, we define
\be
  \Delta_{\rm max} = \max(\Delta M(t)) - \min(\Delta M(t))
\ee
to measure the overall change in any difference curve.
Figure \ref{fig:delmaxPDF} shows the probability distribution
for $\Delta_{\rm max}$.\footnote{We have combined all stellar
population parameters and configurations.  While this does wash
out detailed information that may distinguish one set of stellar
population parameters from another, for the moment we only seek
to understand how likely it is that microlensing fluctuations
will be seen in a lensed SN.  We have examined $\Delta_{\rm max}$
distributions for different stellar populations, and they are
all qualitatitively similar to the distribution shown in
Figure \ref{fig:delmaxPDF}.}  There are two crucial points.
First, the $\Delta_{\rm max}$ distribution is not peaked at
zero, so microlensing fluctuations are more likely than not to
be observed in SN light curves.  Second, the total variation in
a given SN light curve due to microlensing can be quite large.
The $\Delta_{\rm max}$ distribution is peaked at $\sim 0.6$ mag
with a long tail that extends to $> 2$ mag.  Strikingly, the
probability that $\Delta_{\rm max}$ is less than $\sim$0.1 mag is
quite low.  From the cumulative distribution, we estimate that
70\% of lensed SN light curves will have microlensing fluctuations
greater than 0.5 mag, and 25\% will have fluctuations greater
than 1 mag.  We have checked that similar conclusions hold for
other typical values of $\kappa$ and $\gamma$ (specifically, those
used in \refsec{dt}), including both positive and negative parity
situations.

The implication is that microlensing of lensed SNe is almost a
certainty given reasonable stellar populations.

\section{Analysis}

\subsection{Extracting parameters of the stellar population}
\label{sec:extract}

\begin{figure*}
\begin{center}
\includegraphics[width=5.5in]{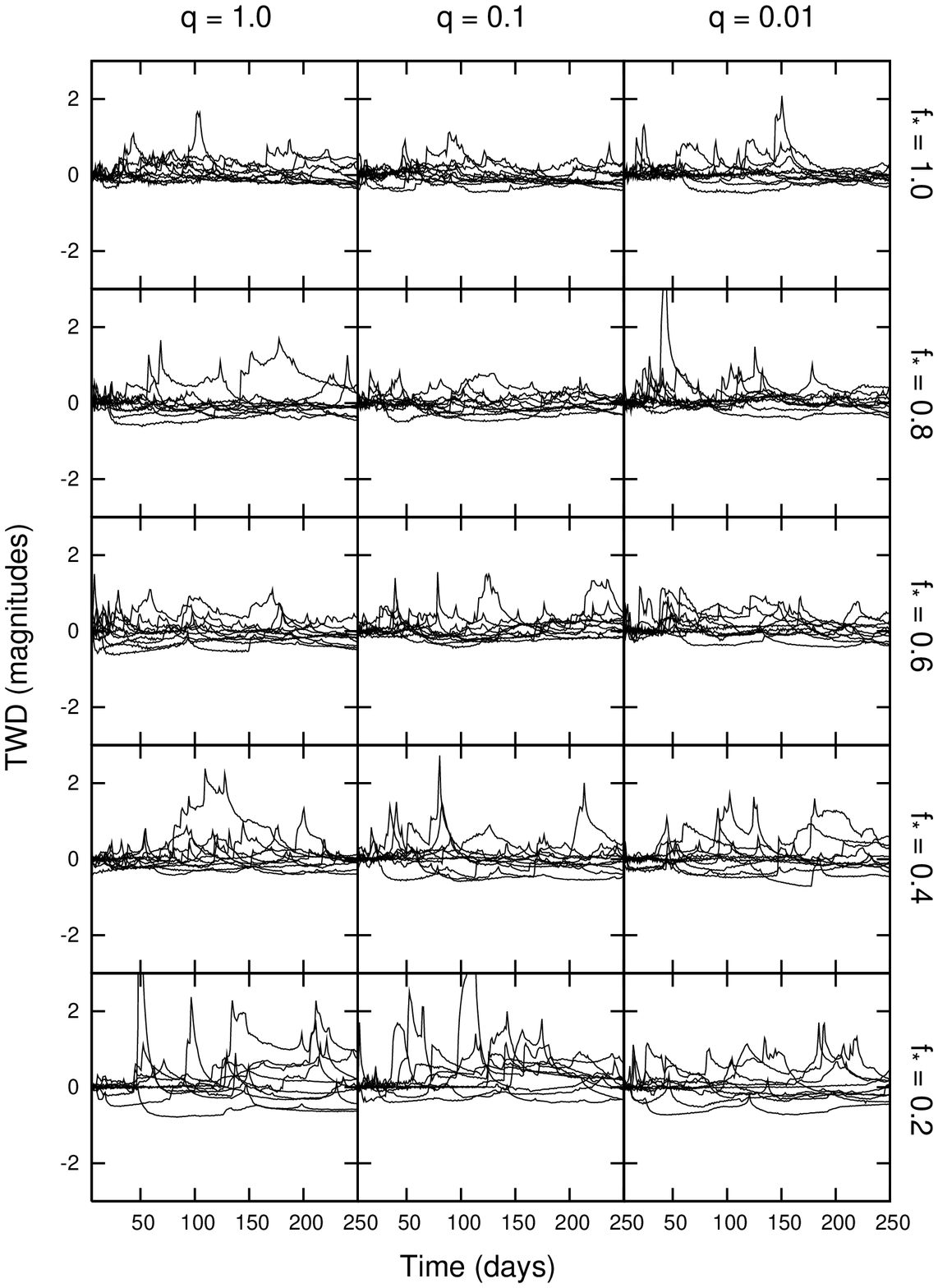}
\end{center}
\caption{
Time-weighted derivatives (TWD) of the difference light curves shown
in Figure \ref{fig:fullsample}.  In general, for fixed $f_*$ the
TWD is more tightly bound about zero for smaller values of $q$.
This effect is more pronounced for fixed $q$ and larger values of
$f_*$.
}\label{fig:fullsampleTWD}
\end{figure*}

Comparing the various panels of Figure \ref{fig:fullsample}, the
eye picks out some qualitative features.  For example, in the
$q=0.01$ cases there seem to be more regions where the difference
light curves are fairly flat than in the $q=1.0$ cases.  To quantify
this feature, we compute the time-weighted derivative (TWD) of the
difference light curves,
\be
  \TWD = -t \times \frac{\partial}{\partial t} \Delta M,
\ee
where the negative sign ensures that TWD is positive for increasing
brightness (i.e., decreasing magnitude).  It is useful to explain
the motivation further.  When the mass range is broad ($q$ is
small), there are a few high-mass stars and they tend to be well
separated.  When the photosphere encounters one of these stars
there can be an abrupt change in the lensing magnification which
leads to a ``step'' in the light curve.  Conversely, between
these abrupt events the magnification fluctuations are small.
A similar argument holds when the stars are diluted by a smooth
mass component ($f_*$ decreases).  While different stellar
configurations certainly produce different light curves, in
general the step-type fluctuations are more characteristic of
low $q$ or low $f_*$.  The time-weighting of the derivative
accounts for finite source effects: the change in brightness
will happen over a longer time scale when a larger source comes
into contact with a stellar caustic.

The TWD curves are shown in Figure \ref{fig:fullsampleTWD}.
There is striking contrast among the different $f_*$ and $q$
combinations.  In general, the $q = 0.01$ cases do seem more
tightly bound about $\TWD = 0$ (i.e., flat difference light
curves); this effect is particularly notable for $f_* = 0.2$.  
There is an even more pronounced effect when comparing 
different $f_*$ values at fixed $q$.  Diluting the stellar
component with smooth dark matter clearly \emph{increases} the
scatter about $\TWD = 0$.

Notice that the TWD curves show large positive spikes, but no
corresponding negative spikes.  This is because there is a
simple lower bound on the TWD, which can be seen as follows.
The difference light curve may be written as
\be \label{eq:DeltaM}
  \Delta M = -2.5 \log \left[\frac{A_I}{\mu_0 \pi \Rsrc^2}\right] ,
\ee
where $A_I$ is the area of the images.  Therefore the TWD is
\be \label{eq:TWD}
  \TWD = \frac{2.5\,t}{\ln(10)} \left[
      \frac{1}{A_I}\,\frac{\partial A_I}{\partial t}
    - \frac{2}{\Rsrc}\,\frac{\partial \Rsrc}{\partial t}
    \right] .
\ee
For linear expansion $\Rsrc = v\,t$ so the second term
simplifies and we obtain
\be
  \TWD = \frac{2.5\,t}{A_I \ln(10)}\frac{\partial A_I}{\partial t} - 
  \frac{5}{\ln(10)}.
\ee
In the first term, it is important to understand that
$\partial A_I / \partial t \geq 0$.  This is because as the
source expands and crosses more and more caustics, images can
never disappear.  Distinct images may merge, but they cannot
vanish because, if a particular source position is covered by
the photosphere at one time, it is covered at all later times as
well.\footnote{The surface brightness of the photosphere may
change with time, but that does not affect our argument because
in equations (\ref{eq:DeltaM}) and (\ref{eq:TWD}) we only need
to specify the \emph{area} of the images; the source surface
brightness has factored out.}  This means that the first term
is non-negative, so we have $\TWD \ge -5/\ln(10) = -2.17$.
In practice, our TWD curves do not even dip below
$\TWD \approx -1$, which suggests that there may be some
other effect bounding the TWD.  Nevertheless, this argument
does explain why there are no large negative spikes, only large
positive spikes.

\begin{figure}
\begin{center}
\includegraphics[width=3.2in]{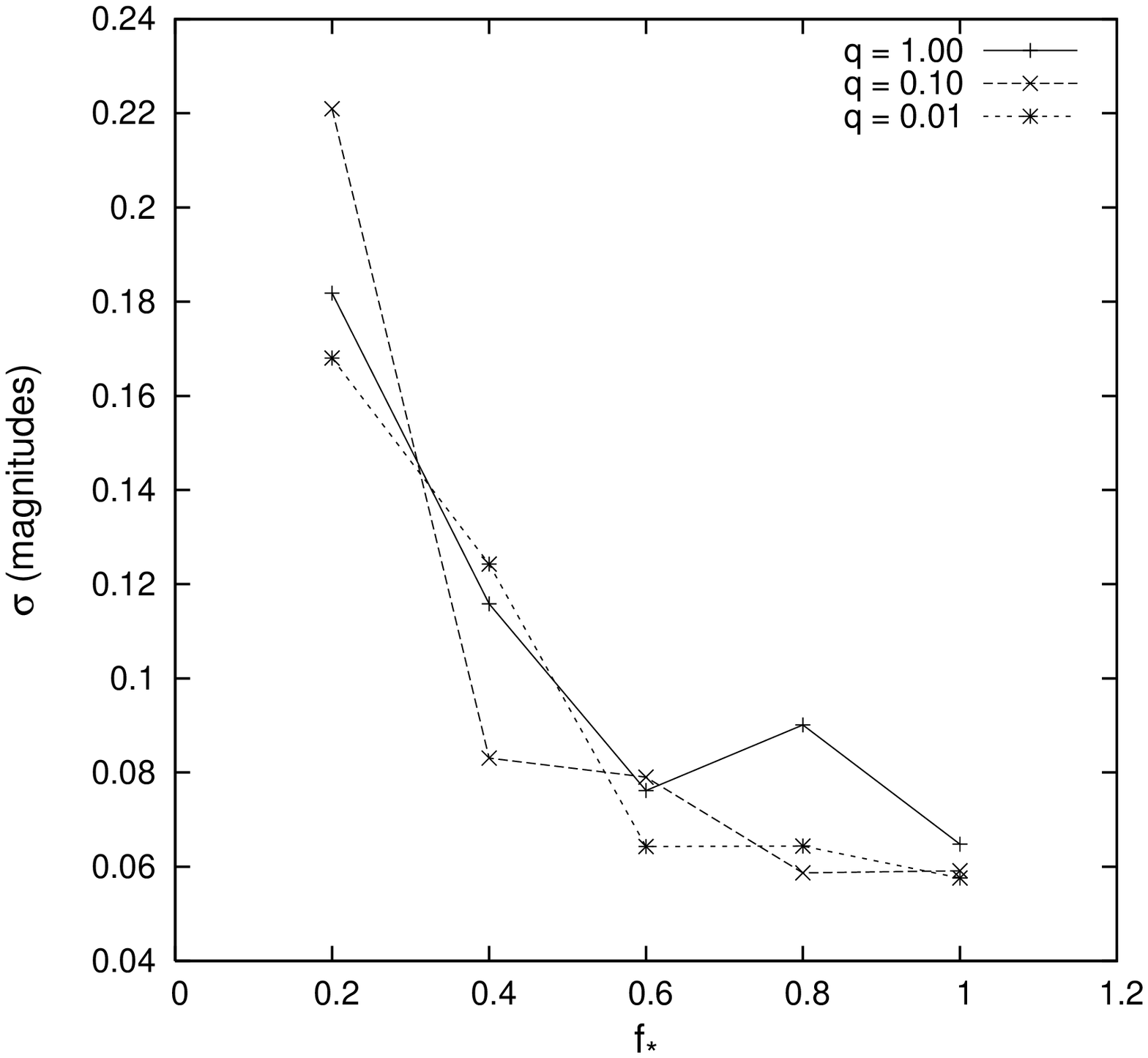}
\end{center}
\caption{
Dispersion $\sigma$ in the TWD distributions, plotted versus $f_*$
for the three values of $q$.  There is a clear decrease of $\sigma$
as $f_*$ is increased from 0.2 to 1.0.  Thus, microlensing fluctuations
in SN light curves may be useful in extracting the relative amounts
of stellar and smooth dark matter at the positions of the lensed
images.
}\label{fig:sigfstar}
\end{figure}

To quantify our sense that different stellar populations have
TWD curves that are more or less tightly bound around zero, we
compute the dispersion in the TWD distribution averaged over
all times and all curves for each panel in
Figure \ref{fig:fullsampleTWD}.  Figure \ref{fig:sigfstar} shows
how the TWD dispersion depends on $f_*$ for the various values
of $q$.  We find a clear decrease in the TWD dispersion as $f_*$
increases.  There is no obvious trend with $q$.  We conclude
that the TWD contains key information about the relative amounts
of stellar versus smooth dark matter mass densities.  Constraints
on the stellar mass function will be more difficult to extract
from the TWD, but perhaps other statistical measures can be found
that will reveal information about the mass function.

\subsection{Time delay uncertainties}
\label{sec:dt}

One goal for lensed SNe is to measure lensing time delays very
precisely.  We now consider whether microlensing may make it
difficult to ``match up'' light curves to determine the time
delay.  To consider a realistic lensing scenario, we consider
the SLACS lens SDSS J2300+0022 and use the lens model by
\citet{SLACS3}.  We place a SN near the center of the source
and solve the lens equation to find the local convergence and
shear at the two SN image positions ($\kappa = \gamma = 0.39$ 
for the positive parity image and $\kappa = \gamma = 0.52$ for 
the negative parity image), and also the time delay between the
images ($\Delta t_0 = 16.5$ days).  We then generate
mock light curves starting from two observed cases: SN 2004ej
and SN 2004ex from the CSP.  We seek to focus on the uncertainties
induced by microlensing, so we limit the uncertainties due to
finite sampling by artificially boosting the time sampling to
$\sim$4 hours with linear interpolation.  We generate
microlensing difference curves using $f_{*}=0.5$
(see \refsec{estfstar}), a uniform mass function ($q=1.0$),
and then add them to the intrinsic light curves to obtain
different realizations of lensed SN light curves.  We create
20 random realizations for each image, which can be used to
construct 400 distinct ``observations.''

For each ``observation,'' we then attempt to recover the time
delay by matching up the light curves from the two images.
Specifically, we minimize the goodness of fit
\be
  \chi^2(\Delta t) = \frac{1}{N} \sum_{i=1}^{N}
    \frac{1}{\sigma_i^2} \left[ D^+(t_i) - D^-(t_i-\Delta t) - k \right]^2 .
\ee
Here $D^+$ and $D^-$ are the positive and negative parity SN
light curves (in magnitudes), respectively.  The normalization
constant $k$ is required because the two images have different 
\emph{macro}-magnifications from the main lens galaxy; in the
absence of microlensing, $k = -2.5 \log(\mu_+/\mu_-)$.  Also,
$N$ is the number of data points for which the shifted light
curves overlap.  We normalize $\chi^2$ by $N$ in order to avoid
a scenario in which the optimization routine takes $\chi^2$ to
zero simply by shifting the light curves so they do not overlap
and there are no terms in the sum.  For simplicitly, we assume
that the photometric uncertainties $\sigma_i$ are Gaussian and
constant in time.

\begin{figure*}
\centerline{
  \includegraphics[width=2.0in]{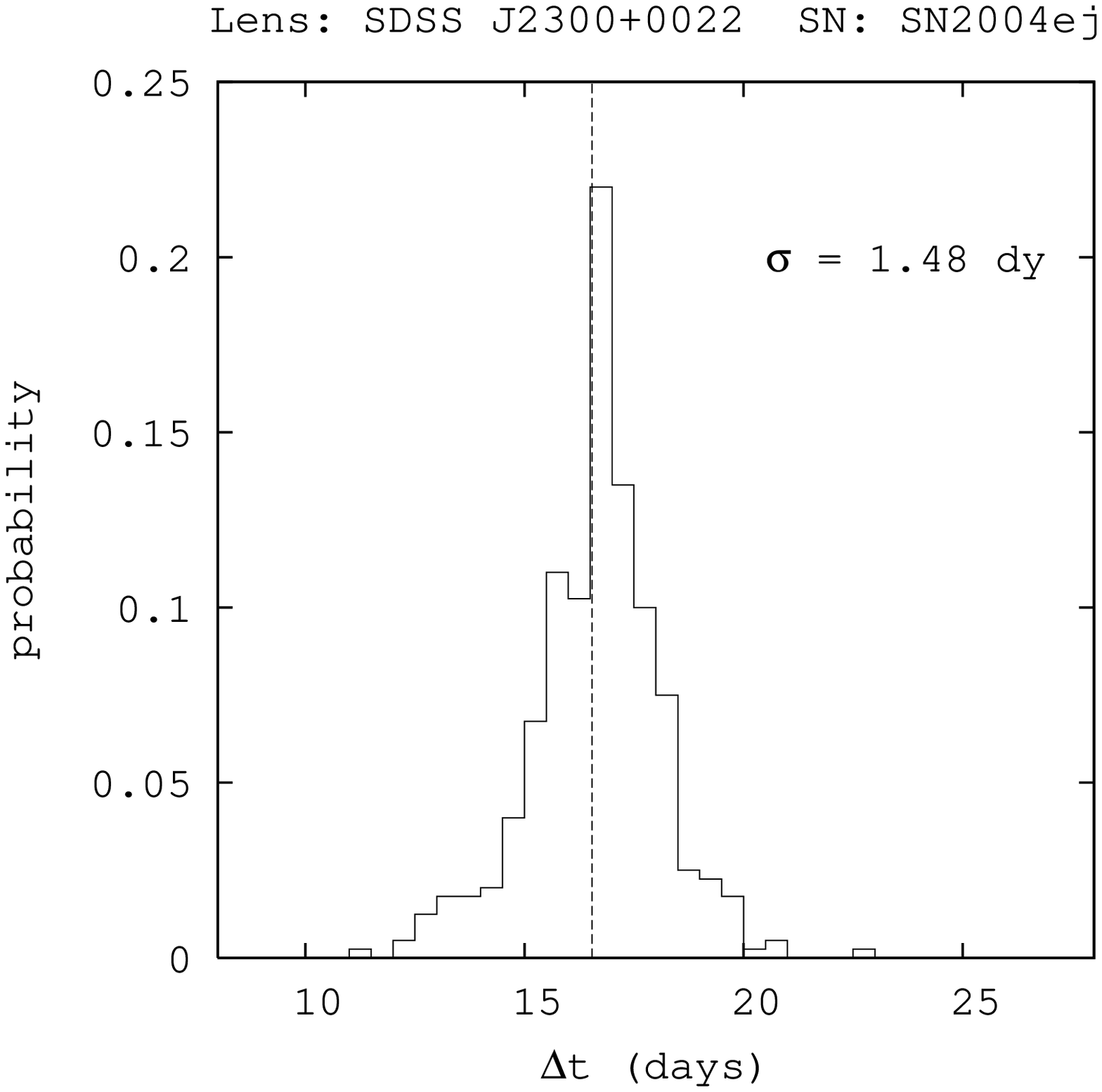}
  \includegraphics[width=2.0in]{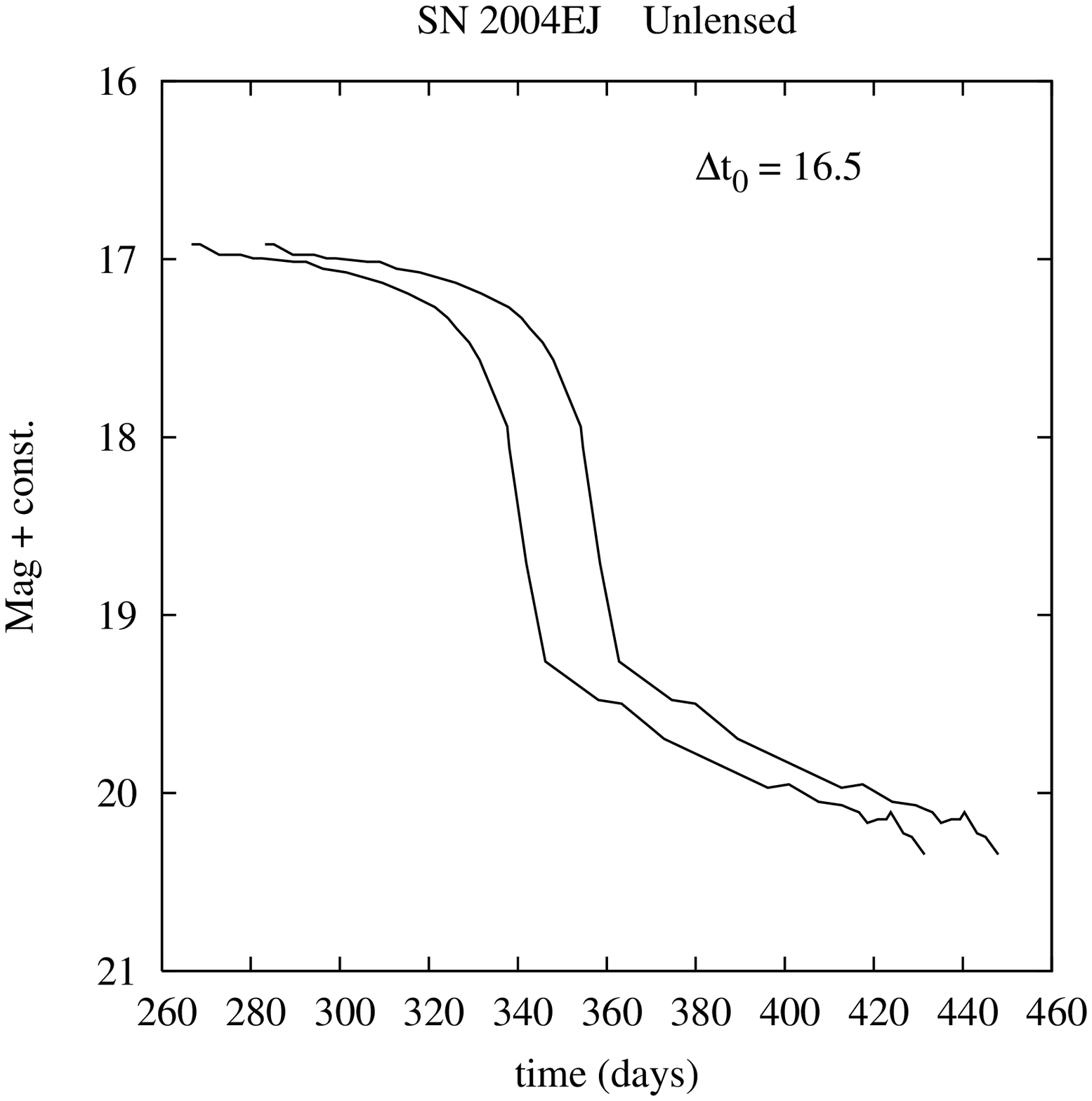}
  \includegraphics[width=2.0in]{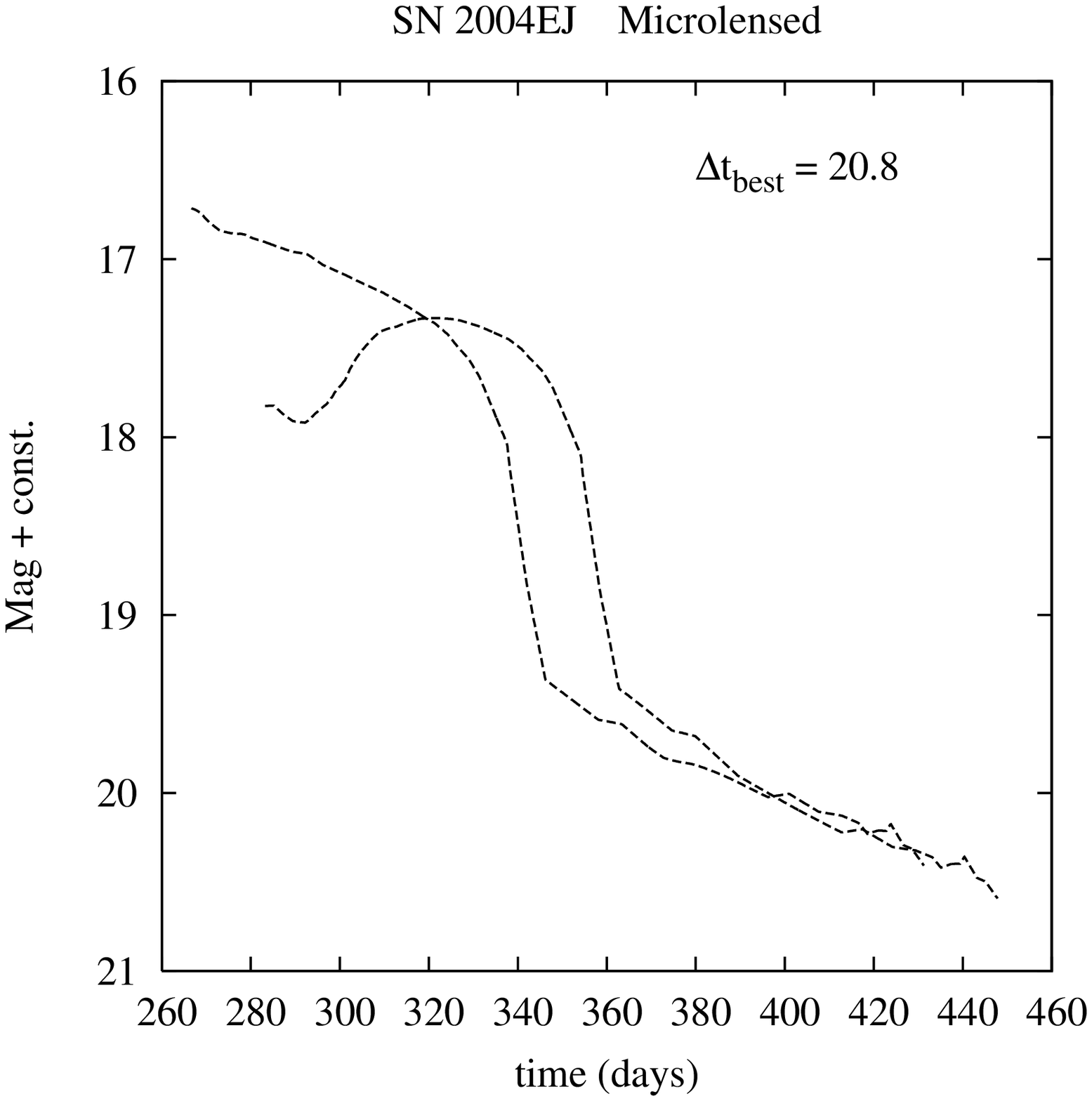}
}
\centerline{
  \includegraphics[width=2.0in]{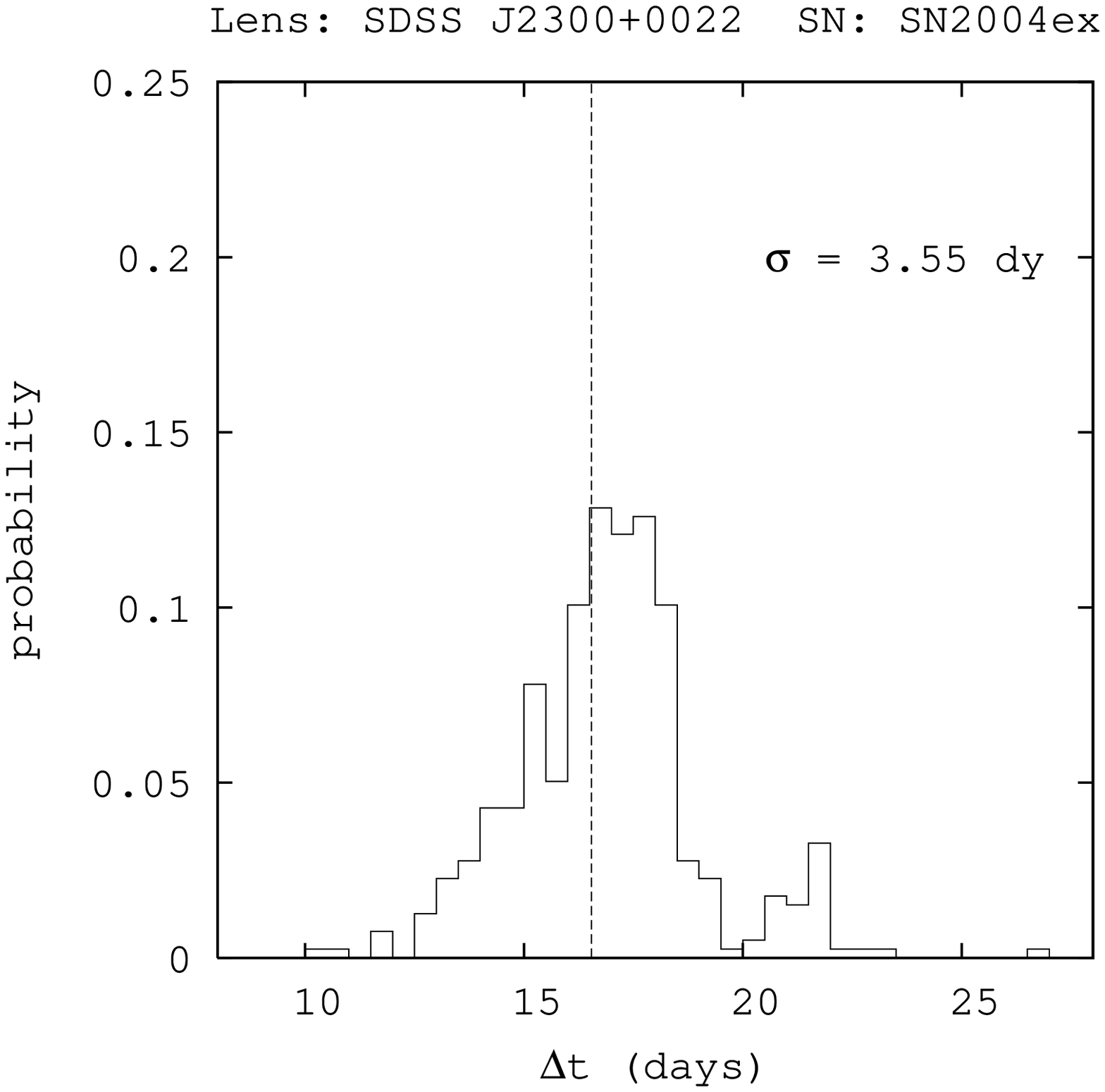}
  \includegraphics[width=2.0in]{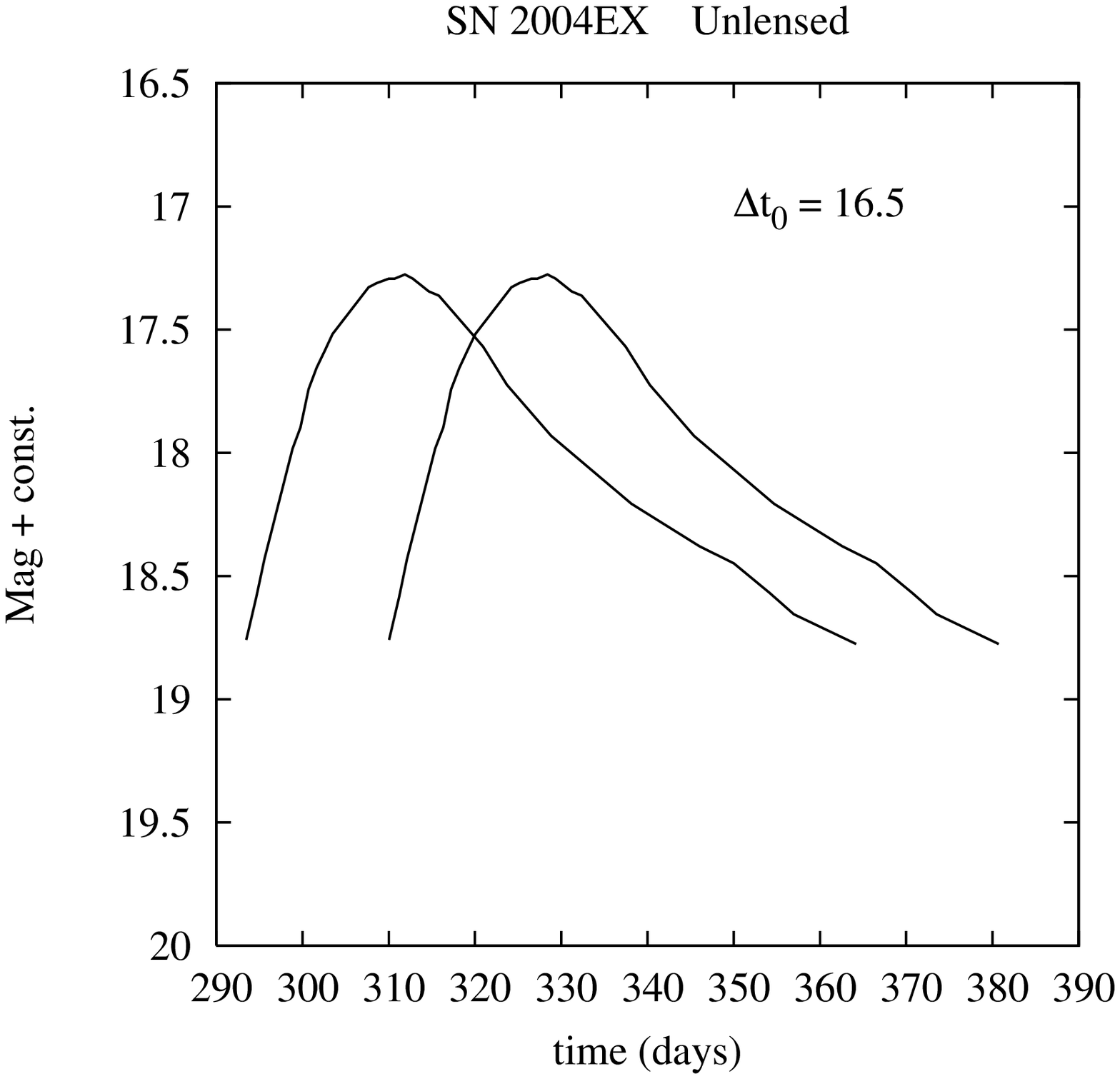}
  \includegraphics[width=2.0in]{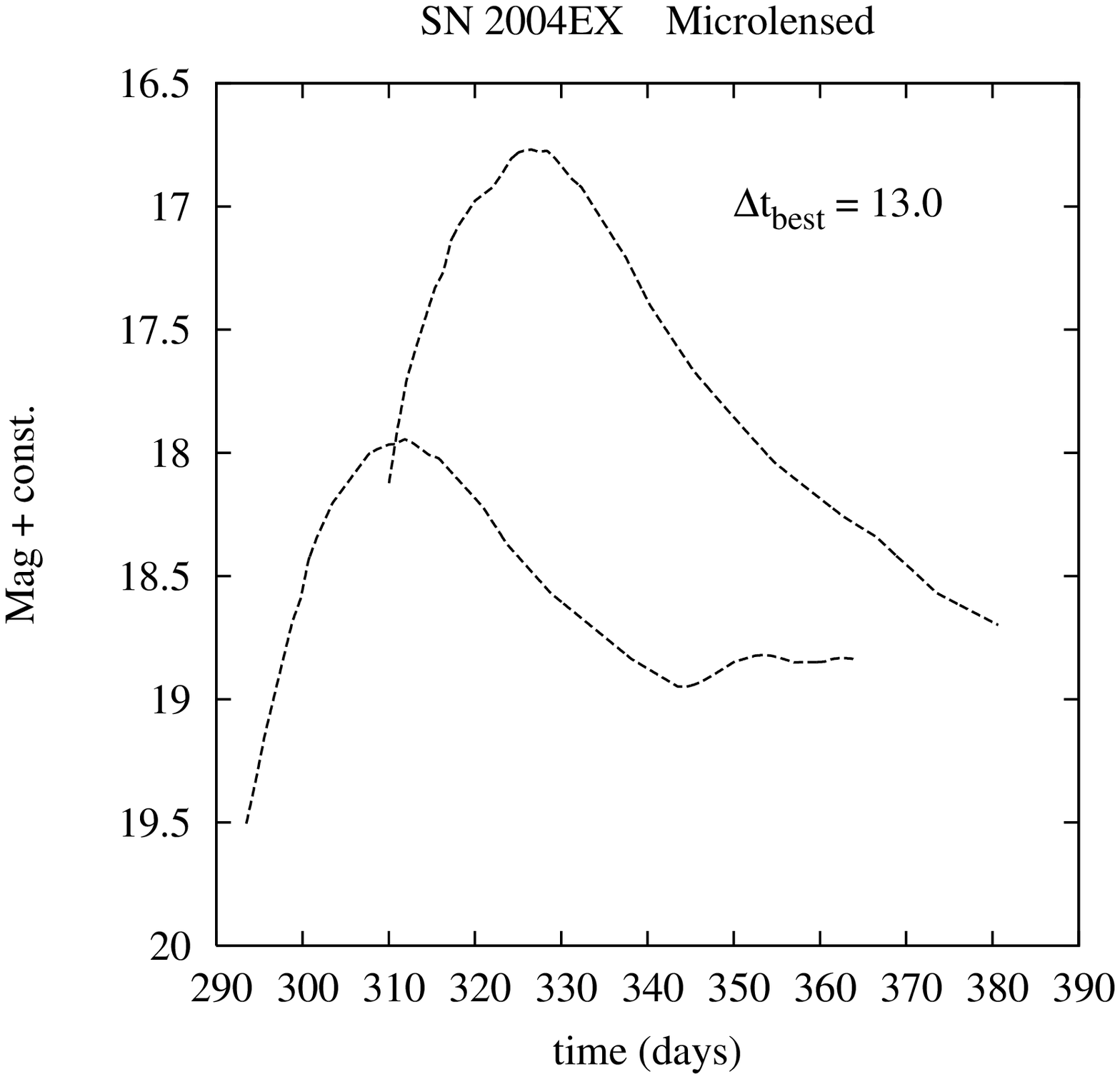}
}
\caption{
Illustration of the effects of microlensing on time delay
measurements.  We create a mock lensed SN using two different
intrinsic SN light curves (SN 2004ej and SN 2004ex, both from
the CSP) and lens/source parameters from the SLACS lens 
J2300+0022 \citep{SLACS3}.
Left column: Histograms of the recovered time delays for 400 mock
observations.  The true time delay $\Delta t_0 = 16.54$ days is
marked with a vertical line.  The histograms are broad (dispersion
$\sigma = 1.48$ days and 3.55 days for SN 2004ej and SN 2004ex 
respectively) and skewed with peaks offset from $\Delta t_0$.  The 
implication is that microlensing effects will render high precision
time delays difficult to measure from multiply imaged SNe.  
Center column: The intrinsic light curves for the two SNe.  The
leading and trailing light curves correspond to the positive and
negative parity images, respectively.
Right column: The positive and negative parity light curves,
including microlensing fluctuations, for one sample mock
observation.
}\label{fig:dthist}
\end{figure*}

Figure \ref{fig:dthist} shows sample observations, together
with histograms of the recovered time delays.  The time delay
histograms are broad: the dispersions are $\sigma = 1.5$ days
for SN 2004ej and 3.6 days for SN 2004ex, implying that
microlensing will make it difficult to measure time delays more
precise than a few days.  Furthermore, the $\Delta t$ histograms
are peaked not at the true time delay but rather at a higher
value, and are significantly skewed.

The differences between the results for the two different
intrinsic SNe can be understood by comparing the intrinsic
light curves.  SN 2004ej has a very distinct, large amplitude
intrinsic feature at $t \approx 340$ days.  The light curve 
decreases rapidly by $\sim$2 mag, which is larger than typical
microlensing effects (see Fig.~\ref{fig:fullsample}).  
In this case, microlensing can perturb
the time delay analysis but cannot fundamentally disrupt the
ability to line up the light curves and measure the time delay.
Conversely, SN 2004ex has intrinsic variability over longer
time scales and, more importantly, of smaller amplitude.
Microlensing introduces fluctuations with similar time scales
and amplitudes, which makes it considerably more difficult to
match up the light curves of the two images.

We consider whether it is possible to improve the time delay
measurements by ``masking out'' portions of the light curve
when the source is more susceptible to microlensing.
Specifically, we repeat the time delay analysis while ignoring
the first 25\% of the light curve when the source is small and
microlensing fluctuations are largest, and the last 25\% when
the source is becoming faint.  For SN 2004ej this effectively
isolates the large intrinsic drop around $t \approx 340$ days,
and leads to a smaller time delay dispersion of $\sigma = 0.6$
days.  In other words, if the light curve has a strong intrinsic
feature, it can pay to focus on this feature for measuring the
time delay.  For SN 2004 ex, by contrast, masking the first and
last parts of the light curve leads to a larger dispersion of
$\sigma = 4.40$ days.  This is not surprising since, with no
dominant intrinsic feature, masking out part of the data only
serves to decrease the number of data points.

In other words, the extent to which microlensing corrupts time
delay measurements will depend on whether there are strong
intrinsic features in the SN light curve.  Of course, for a
real lensed SN the intrinsic light curve will not be known,
and the challenge will be to determine which observed features
are intrinsic and which are due to microlensing effects.
Nevertheless, we can conclude in general that microlensing
effects must be taken into account when attempting to extract
high precision time delays from multiply imaged SNe.

\section{Conclusions}

In anticipation of the discovery of lensed SNe in the near
future with SLACS and SNAP, we have undertaken a pilot study
of the fluctuations in the observed light curves due to
microlensing by stars in the lens galaxy.  Our main findings
can be summarized as follows.
\begin{itemize}

\item
The microlensing optical depth and the angular size of the
source SN virtually ensure that microlensing fluctuations will
be present in lensed SN light curves.

\item
Microlensing can produce an overall brightness offset of 
$\sim$1 mag, short time scale brightness fluctuations of
$\sim$0.1--0.2 mag, and longer time scale fluctuations of
$\sim$0.5--1.0 mag.

\item
We estimate that $\sim$70\% of lensed SN light curves will
have microlensing fluctuations larger than 0.5 mag, and
$\sim$25\% will have fluctuations larger than 1 mag.  The
probability that microlensing effects will \emph{not} be
present in any given light curve is $\la 5\%$.

\item
Microlensing fluctuations can generate dramatic and easily 
observable changes to the intrinsic SN light curve.  These
features will complicate efforts to use time delays to measure
the Hubble constant to high precision.  Without detailed knowledge 
of the intrinsic structure of the light curve, time delays 
accurate to better than a few days will be difficult to obtain.

\end{itemize}

We should say that if microlensing fluctuations are \emph{not}
observed in lensed SN light curves, there are two possible
interpretations: either the ratio of stellar mass density to
smooth dark matter mass density (at the image positions) is so
low that the microlensing optical depth is near zero; or the SN 
photosphere undergoes an initial period of very rapid expansion
($v \gg 15,000$ km/s at early times).

The microlensing fluctuations depend on the stellar mass fraction
and mass function in the lens galaxy, although there are such
large difference between different random star configurations
that it may be challenging to extract the stellar population
parameters.  We have introduced the time-weighted derivative (TWD)
of the light curve as a quantity that may be useful for this
analysis.  The TWD is relevant because of the correlation between
fluctuation time scales, stellar Einstein radii, and the
photosphere expansion velocity.  We find that TWD statistics
may be most useful for determining the relative abundances of
stars and dark matter at the image positions.  More generally,
the amount of structure in the light curves is a sign that more
effort should be exerted to figure out how to analyze lensed SN
light curves to determine properties of the stellar population.

Once a lensed SN is typed via spectroscopy or the late-time shape
of the photometric light curve, it may be possible to model the
intrinsic light curve.  The efficacy of EPM implies that the
source geometry at different epochs in the observations will be
fairly well known, and so a light curve fitting method
\citep[such as that described by][]{csk} may ultimately be the 
best analysis.  The fitting method must be extended to efficiently
incorporate \emph{expanding} sources as opposed to \emph{moving}
sources; however the basic idea of finding a family of possible
stellar configurations and using a Bayesian likelihood analysis 
to extract mass function parameters will remain the same.

\acknowledgements
We thank Mark Phillips and the Carnegie Supernova Project for
providing data for SN 2004ej and SN 2004ex.  We thank
Gary Bernstein and the anonymous referee
for helpful comments and suggestions.  CRK thanks
Mike Gladders for stimulating discussions at the conception of
this project.  This project is supported by grant HST-AR-10668
from the Space Telescope Science Institute, which is operated
by the Association of Universities for Research in Astronomy,
Inc., under NASA contract NAS5-26555.

\end{document}